\documentclass[aps,prd,amsmath,two column,amssymb,showpacs]{revtex4}
\usepackage{txfonts}
\usepackage{mathbbold}
\usepackage{epsfig,bm,dcolumn}
\usepackage{graphicx}
\usepackage{color}
\usepackage{overpic}

\newcommand{\feyn}[1]{\setbox0=\hbox{\ensuremath{#1}}\hbox to\wd0{\hbox to0pt{\hbox to\wd0{\hss/\hss}\hss}\box0}}

\begin{document}

\title{Solid-state calculation of crystalline color superconductivity}

\author{Gaoqing Cao,$^{1}$ Lianyi He,$^{2}$ and Pengfei Zhuang$^{1}$}

\affiliation{1 Department of Physics, Tsinghua University and Collaborative Innovation Center of Quantum Matter, Beijing 100084, China\\
2 Theoretical Division, Los Alamos National Laboratory, Los Alamos, New Mexico 87545, USA}

\date{\today}

\begin{abstract}
It is generally believed that the inhomogeneous Larkin-Ovchinnikov-Fulde-Ferrell (LOFF) phase appears in a color superconductor when the pairing between different quark flavors is under the circumstances of mismatched Fermi surfaces. However, the real crystal structure of the LOFF phase is still unclear because an exact treatment of 3D crystal structures is rather difficult. In this work we present a solid-state-like calculation of the ground-state energy of the body-centered cubic (BCC) structure for two-flavor pairing by diagonalizing the Hamiltonian matrix in the Bloch space without assuming a small amplitude of the order parameter. We develop a computational scheme to overcome the difficulties in diagonalizing huge matrices. Our results show that the BCC structure is energetically more favorable than the 1D modulation in a narrow window around the conventional LOFF-normal phase transition point, which indicates the significance of the higher-order terms in the Ginzburg-Landau approach.
\end{abstract}


\maketitle

\section{Introduction}

The ground state of exotic fermion Cooper pairing with mismatched Fermi surfaces is a longstanding problem in the theory of superconductivity
~\cite{Casalbuoni2004}. In electronic superconductors, the mismatched Fermi surfaces are normally induced by the Zeeman energy splitting $2\delta\mu$ in a magnetic field. For $s$-wave pairing at weak coupling, it is known that, at a critical field $\delta\mu_1=0.707\Delta_0$
where $\Delta_0$ is the pairing gap at vanishing mismatch, a first-order phase transition from the gapped BCS state to the normal state occurs
~\cite{CC1962}. Further theoretical studies showed that the inhomogeneous Larkin-Ovchinnikov-Fulde-Ferrell (LOFF) state can survive in a narrow window $\delta\mu_1<\delta\mu<\delta\mu_2$, where the upper critical field $\delta\mu_2=0.754\Delta_0$ \cite{LO1964,FF1964}. However, since the thermodynamic critical field is much lower than $\delta\mu_1$ due to strong orbit effect, it is rather hard to observe the LOFF state in ordinary superconductors \cite{CC1962}. In recent years, experimental evidences for the LOFF state in some superconducting materials have been
reported~\cite{Heavyfermion,HighTc,Organic,FeSe}.

On the other hand, exotic pairing phases have promoted new interest in the studies of dense quark matter under the circumstances of compact stars
\cite{Alford2001,Bowers2002,Shovkovy2003,Alford2004,EFG2004,Huang2004,Casalbuoni2005,Fukushima2005,Ren2005,Gorbar2006,Anglani2014} and ultracold atomic Fermi gases with population imbalance \cite{Atomexp,Sheehy2006}. Color superconductivity in dense quark matter appears due to the attractive interactions in certain diquark channels~\cite{CSC01,CSC02,CSC03,CSC04,CSCreview}. Because of the constraints from Beta equilibrium and electric charge neutrality, different quark flavors ($u$, $d$, and $s$) acquire mismatched Fermi surfaces. Quark color superconductors under compact-star constraints as well as atomic Fermi gases with population imbalance therefore provide rather clean systems to realize the long-sought exotic LOFF phase.

Around the tricritical point in the temperature-mismatch phase diagram, the LOFF phase can be studied rigorously by using the Ginzburg-Laudau (GL) analysis since both the gap parameter and the pair momentum are vanishingly small \cite{Casalbuoni2004}. It was found that the solution with two antipodal wave vectors is the preferred one \cite{Buzdin1997,Combescot2002,Ye2007}. However, the real ground state of the LOFF phase is still debated due to the limited theoretical approaches at zero temperature. So far rigorous studies of the LOFF phase at zero temperature are restricted to its 1D structures including the Fulde-Ferrell (FF) state with a plane-wave form $\Delta(z)=\Delta e^{2iqz}$ and the Larkin-Ovhinnikov (LO) state with an antipodal-wave form $\Delta(z)=2\Delta \cos(2qz)$. A recent self-consistent treatment of the 1D modulation \cite{Buballa2009} show that a solitonic lattice is formed near the lower critical field, and the phase transition to the BCS state is continuous. Near the upper critical field the gap function becomes sinusoidal, and the transition to the normal state is of first order.

In addition to these 1D structures, there exist a large number of 3D crystal structures. The general form of a crystal structure of the order parameter can be expressed as
\begin{eqnarray}\label{crystal}
\Delta({\bf r})=\sum_{k=1}^P\Delta e^{2iq\hat{\bf n}_k\cdot{\bf r}}.
\end{eqnarray}
A specific crystal structure corresponds to a multi-wave configuration determined by the $P$ unit vectors ${\bf n}_{k}$ ($k=1,2,...,P$). In general, we expect two competing mechanisms: Increasing the number of waves tends to lower the energy, but it may also causes higher repulsive interaction energy between different wave directions.
In a pioneer work, Bowers and Rajagopal investigated 23 different crystal structures by using the GL approach~\cite{Bowers2002}, where the grand potential measured with respect to the normal state was expanded up to the order $O(\Delta^6)$,
\begin{eqnarray}\label{GL}
\frac{\delta\Omega(\Delta)}{{\cal N}_{\rm F}}= P\alpha\Delta^2+\frac{1}{2}\beta\Delta^4+\frac{1}{3}\gamma\Delta^6+O(\Delta^8)
\end{eqnarray}
with ${\cal N}_{\rm F}$ being the density of state at the Fermi surface and the pair momentum fixed at the optimal value $q=1.1997\delta\mu$. Among the structures with $\gamma>0$, the favored one seems to be the body-centered cubic (BCC) with $P=6$ \cite{BCC}. Further, it was conjectured that the face-centered cubic (FCC) with $P=8$ \cite{FCC} is the preferred structure since its $\gamma$ is negative and the largest~\cite{Bowers2002}. For BCC structure, the GL analysis up to the order $O(\Delta^6)$ predicts a strong first-order phase transition at $\delta\mu_*\simeq3.6\Delta_0$ with the gap parameter $\Delta\simeq0.8\Delta_0$~\cite{Bowers2002}. The prediction of a strong first-order phase transition may invalidate the GL approach itself. On the other hand, by using the quasiclassical equation approach with a Fourier expansion for the order parameter, Combescot and Mora~\cite{Combescot2004,Combescot2005} predicted that the BCC-normal transition is of rather weak first order: The upper critical field $\delta\mu_*$ is only about $4\%$ higher than $\delta\mu_2$ with $\Delta\simeq0.1\Delta_0$ at $\delta\mu=\delta\mu_*$. If this result is reliable, it indicates that the higher-order expansions in the GL analysis is important for quantitative predictions. To understand this intuitively, let us simply add the eighth-order term $\frac{\eta}{4}\Delta^8$ to the GL potential (\ref{GL}). A detailed analysis of the influence of a positive $\eta$ on the phase transition is presented in Appendix A. We find that with increasing $\eta$, the first-order phase transition becomes weaker and the upper critical field $\delta\mu_*$ decreases. For $\eta\rightarrow+\infty$, the phase transition approaches second order and $\delta\mu_*\rightarrow\delta\mu_2$. Therefore, to give more precise predictions we need to study the higher-order expansions and the convergence property of the GL series, or use a different way to evaluate the grand potential without assuming a small value of $\Delta$.

For a specific crystal structure given by (\ref{crystal}), it is periodic in coordinate space. As a result, the eigenvalue equation for the fermionic excitation spectrum in this periodic pair potential, which is known as the Bogoliubov-de Gennes (BdG) equation, is in analogy to the Schr\"{o}dinger  equation of quantum particles in a periodic potential. This indicates that the fermionic excitation spectrum has a band structure, which can be solved from the BdG equation. The grand potential can be directly evaluated once the fermionic excitation spectrum is known \cite{Buballa2009}.
In this work, we present a solid-state-like calculation of the grand potential of the BCC structure. Our numerical results show that the phase transition from the BCC state to the normal state is of rather weak first order, consistent with the work by Combescot and Mora~\cite{Combescot2004,Combescot2005}. This implies that it is quite necessary to evaluate the higher-order terms in the GL expansion to improve the quantitative predictions.

\section{Thermodynamic Potential}

To be specific, we consider a general effective Lagrangian for two-flavor quark pairing at high density and at weak coupling. The Lagrangian density is given by~\cite{Casalbuoni2004}
\begin{equation}
{\cal L}=\psi^\dagger[i\partial_t-\varepsilon(\hat{\bf p})+\hat{\mu}]\psi+{\cal L}_{\rm int},
\end{equation}
where $\psi=(\psi_{\rm u},\psi_{\rm d})^{\rm T}$ denotes the two-flavor quark field and $\varepsilon(\hat{\bf p})$ is the quark dispersion with the momentum operator $\hat{\bf p}=-i\mbox{\boldmath{$\nabla$}}$. In the momentum representation we have $\varepsilon({\bf p})=|{\bf p}|$. The quark chemical potentials are specified by the diagonal matrix $\hat{\mu}={\rm diag}(\mu_{\rm u},\mu_{\rm d})$ in the flavor space, where
\begin{eqnarray}
\mu_{\rm u}=\mu+\delta\mu,\ \ \ \ \
\mu_{\rm d}=\mu-\delta\mu.
\end{eqnarray}
The interaction Lagrangian density which leads to Cooper pairing between different flavors can be expressed as~\cite{Casalbuoni2004}
\begin{eqnarray}
{\cal L}_{\rm{int}}=g(\psi^\dagger\sigma_2\psi^*)(\psi^{\rm T}\sigma_2\psi),
\end{eqnarray}
where $g$ is the coupling constant and $\sigma_2$ is the second Pauli matrix in the flavor space. Notice that we have neglected the antiquark degree of freedom because it plays no role at high density and at weak coupling. We have also neglected the color and spin degrees of freedom, which simply
give rise to a degenerate factor.

Color superconductivity is characterized by nonzero expectation value of the diquark field $\varphi(t,{\bf r})=-2ig\psi^{\rm T}\sigma_2\psi$.
For the purpose of studying inhomogeneous phases, we set the expectation value of $\varphi(t,{\bf r})$ to be static but inhomogeneous, i.e.,
$\langle\varphi(t,{\bf r})\rangle=\Delta({\bf r})$. With the Nambu-Gor'kov spinor $\Psi=(\psi\ \ \psi^*)^{\rm T}$, the mean-field Lagrangian reads
\begin{eqnarray}
{\cal L}_{\rm{MF}}=\frac{1}{2}\Psi^\dagger\left(\begin{array}{cc} i\partial_t-\varepsilon(\hat{\bf p})+\hat{\mu} & -i\sigma_2\Delta({\bf r})\\
i\sigma_2\Delta^*({\bf r})& i\partial_t+\varepsilon(\hat{\bf p})-\hat{\mu} \end{array}\right)\Psi-\frac{|\Delta({\bf r})|^2}{4g}.
\end{eqnarray}
The order parameters of the BCC and FCC structures can be expressed as
\begin{eqnarray}
\Delta({\bf r})=2\Delta\left[\cos\left(2qx\right)
+\cos\left(2qy\right)+\cos\left(2qz\right)\right]
\end{eqnarray}
and
\begin{eqnarray}
\Delta({\bf r})=8\Delta\cos\left(\frac{2qx}{\sqrt{3}}\right)\cos\left(\frac{2qy}{\sqrt{3}}\right)\cos\left(\frac{2qz}{\sqrt{3}}\right),
\end{eqnarray}
respectively. Therefore, we consider a 3D periodic structure where the unit cell is spanned by three linearly independent vectors ${\bf a}_1=a{\bf e}_x$, ${\bf a}_2=a{\bf e}_y$, and ${\bf a}_3=a{\bf e}_z$ with $a=\pi/q$ for BCC and $a=\sqrt{3}\pi/q$ for FCC. The order parameter is periodic in space, i.e., $\Delta({\bf r})=\Delta({\bf r}+{\bf a}_i)$. It can be decomposed into a discrete set of Fourier components,
\begin{eqnarray}
\Delta({\bf r})=\sum_{{\bf G}}\Delta_{\bf G}e^{i{\bf G}\cdot {\bf r}}=\sum_{l,m,n=-\infty}^\infty \Delta_{[lmn]}e^{i{\bf G}_{[lmn]}\cdot {\bf r}},
\end{eqnarray}
where the reciprocal lattice vector ${\bf G}$ reads
\begin{eqnarray}
{\bf G}={\bf G}_{[lmn]}=\frac{2\pi}{a}\left(l{\bf e}_x+m{\bf e}_y+n{\bf e}_z\right),\ \ \ l,m,n\in \mathbb{Z}.
\end{eqnarray}
The Fourier component $\Delta_{\bf G}=\Delta_{[lmn]}$ can be evaluated as
\begin{eqnarray}
&&\Delta_{\bf G}=\Delta\ \Big[\left(\delta_{l,1}+\delta_{l,-1}\right)\delta_{m,0}\delta_{n,0}
+\delta_{l,0}\left(\delta_{m,1}+\delta_{m,-1}\right)\delta_{n,0}\nonumber\\
&&\ \ \ \ \ \ \ \ \ \  +\ \delta_{l,0}\delta_{m,0}\left(\delta_{n,1}+\delta_{n,-1}\right)\Big]
\end{eqnarray}
and
\begin{eqnarray}
\Delta_{\bf G}=\Delta \left(\delta_{l,1}+\delta_{l,-1}\right)\left(\delta_{m,1}+\delta_{m,-1}\right)\left(\delta_{n,1}+\delta_{n,-1}\right)
\end{eqnarray}
for BCC and FCC structures, respectively.

Then we consider a finite system in a cubic box defined as $x,y,z\in[-L/2,L/2]$ with the length $L=Na$. For convenience we impose the periodic boundary condition. The thermodynamic limit is reached by setting $N\rightarrow\infty$. Using the momentum representation, we have the Fourier transformation
\begin{eqnarray}
\Psi(\tau,{\bf r})=V^{-1/2}\sum_{\nu,{\bf p}}\Psi_{\nu{\bf p}}e^{-i(\omega_\nu\tau-{\bf p}\cdot{\bf r})}.
\end{eqnarray}
Here $V$ is the volume of the system, $\omega_\nu=(2\nu+1)\pi T (\nu\in \mathbb{Z})$ is the fermion Matsubara frequency, and the quantized momentum ${\bf p}$ is given by
\begin{equation}
{\bf p}=\frac{2\pi}{L}(l{\bf e}_x+m{\bf e}_y+n{\bf e}_z)
\end{equation}
with $l,m,n\in \mathbb{Z}$. The partition function of the system is given by
\begin{equation}
{\cal Z}=\int [d\Psi][d\Psi^\dagger]e^{-{\cal S}}
\end{equation}
with the Euclidean action ${\cal S}=-\int_0^{1/T}d\tau\int_Vd^3{\bf r}{\cal L}$. The grand potential per volume reads
\begin{equation}
\Omega=-\frac{T}{V}\ln{\cal Z}.
\end{equation}
In the mean-field approximation, the action ${\cal S}$ is quadratic. Therefore, the partition function ${\cal Z}$ and grand potential $\Omega$ can be evaluated. Using the Fourier expansions for $\Psi$ and $\Delta$, we obtain the mean-field  action
\begin{eqnarray}
{\cal S}_{\rm{MF}}=\frac{V}{T}\!\sum_{\bf G}\!\frac{|\Delta_{\bf G}|^2}{4g}-\frac{1}{2T}\!\sum_{\nu,{\bf p},{\bf p}^\prime}\!
\Psi^\dagger_{\nu{\bf p}}
\left(i\omega_\nu\delta_{{\bf p},{\bf p}^\prime} -{\cal H}_{{\bf p},{\bf p}^\prime}\right)\Psi_{\nu{\bf p}^\prime},
\end{eqnarray}
where the effective Hamiltonian matrix ${\cal H}_{{\bf p},{\bf p}^\prime}$ reads
\begin{eqnarray}
{\cal H}_{{\bf p},{\bf p}^\prime}=\left(\begin{array}{cc} (|{\bf p}|-\hat{\mu})\delta_{{\bf p},{\bf p}^\prime}
& i\sigma_2\sum_{\bf G}\Delta_{\bf G}\delta_{{\bf G},{\bf p}-{\bf p}^\prime}
\\ -i\sigma_2\sum_{\bf G}\Delta_{\bf G}^*\delta_{{\bf G},{\bf p}^\prime-{\bf p}}
& -(|{\bf p}|-\hat{\mu})\delta_{{\bf p},{\bf p}^\prime}  \end{array}\right).
\end{eqnarray}
The effective Hamiltonian ${\cal H}_{{\bf p},{\bf p}^\prime}$ is a huge matrix in Nambu-Gor'kov, flavor, and (discrete) momentum spaces. It is Hermitian and can in principle be diagonalized. Assuming that the eigenvalues of ${\cal H}_{{\bf p},{\bf p}^\prime}$ is denoted by $E_\lambda$, we can formally express the grand potential as
\begin{eqnarray}
\Omega=\frac{1}{4g}\sum_{\bf G}|\Delta_{\bf G}|^2-\frac{1}{2V}\sum_\lambda{\cal W}(E_\lambda),
\end{eqnarray}
where the function ${\cal W}(E)=\frac{E}{2}+T\ln(1+e^{-E/T})$. The summation over ${\bf G}$ can be worked out as
$\sum_{\bf G}|\Delta_{\bf G}|^2=P\Delta^2$.

In practice, diagonalization of the matrix ${\cal H}_{{\bf p},{\bf p}^\prime}$ is infeasible. However, ${\cal H}$ can be brought into a block-diagonal form with $N^3$ independent blocks in the momentum space according to the famous Bloch theorem~\cite{Buballa2009}. To understand this, we consider the eigenvalue equation for the fermionic excitation spectrum in the coordinate space, which is known as the BdG equation. For our system, the BdG equation reads
\begin{eqnarray}
\left(\begin{array}{cc} \varepsilon(-i\mbox{\boldmath{$\nabla$}})-\hat{\mu}
& i\sigma_2\Delta({\bf r}) \\ -i\sigma_2\Delta^*({\bf r})
& -\varepsilon(-i\mbox{\boldmath{$\nabla$}})+\hat{\mu}  \end{array}\right)\phi_\lambda({\bf r})
=E_\lambda\phi_\lambda({\bf r}).
\end{eqnarray}
According to the Bloch theorem, the solution of the eigenfunction $\phi_\lambda({\bf r})$ takes the form of the so-called Bloch function. We have
\begin{equation}
\phi_\lambda({\bf r})=e^{i{\bf k}\cdot{\bf r}}\phi_{\lambda{\bf k}}({\bf r}),
\end{equation}
where ${\bf k}$ is the momentum in the Brillouin zone (BZ) and the function $\phi_{\lambda{\bf k}}({\bf r})$ has the same periodicity as the order parameter $\Delta({\bf r})$. We therefore have the similar Fourier expansion
\begin{equation}
\phi_{\lambda{\bf k}}({\bf r})=\sum_{\bf G}\phi_{\bf G}({\bf k})e^{i{\bf G}\cdot {\bf r}}.
\end{equation}
Substituting this expansion into the BdG equation, for a given ${\bf k}$ we obtain a matrix equation
\begin{eqnarray}
\sum_{{\bf G}^\prime}{\cal H}_{{\bf G},{\bf G}^\prime}({\bf k})\phi_{{\bf G}^\prime}({\bf k})=E_\lambda({\bf k}) \phi_{\bf G}({\bf k}),
\end{eqnarray}
where the matrix ${\cal H}_{{\bf G},{\bf G}^\prime}({\bf k})$ is given by
\begin{eqnarray}
\left(\begin{array}{cc} (|{\bf k}+{\bf G}|-\hat{\mu})\delta_{{\bf G},{\bf G}^\prime}
& i\sigma_2\Delta_{{\bf G}-{\bf G}^\prime} \\ -i\sigma_2\Delta^*_{{\bf G}^\prime-{\bf G}}
& -(|{\bf k}+{\bf G}|-\hat{\mu})\delta_{{\bf G},{\bf G}^\prime}  \end{array}\right).
\end{eqnarray}
This shows that, for a given ${\bf k}$-point in the BZ, we can solve the eigenvalue spectrum $\{E_{\lambda}({\bf k})\}$ by diagonalizing the
matrix ${\cal H}_{{\bf G},{\bf G}^\prime}({\bf k})$. Without loss of generality, the BZ can be chosen as
$k_x,k_y,k_z\in[-\pi/a,\pi/a]$. For a quantized volume $V$ containing $N^3$ unit cells, we have $N^3$ allowed
momenta ${\bf k}$ in the BZ. Accordingly, the grand potential is now given by
\begin{eqnarray}
\Omega=\frac{P\Delta^2}{4g}-\frac{1}{2V}\sum_{{\bf k}\in {\rm BZ}}\sum_\lambda{\cal W}[E_\lambda({\bf k})].
\end{eqnarray}
In the thermodynamic limit $N\rightarrow\infty$, the summation $\frac{1}{V}\sum_{{\bf k}\in {\rm BZ}}$ is replaced by an integral over the BZ.

The Hamiltonian matrix ${\cal H}_{{\bf G},{\bf G}^\prime}({\bf k})$ can be further simplified to lower the matrix size. After a proper rearrangement of the eigenvector $\phi_{\bf G}$, we find that ${\cal H}$ can be decomposed into two blocks. We have
${\cal H}={\cal H}_{\Delta,\delta\mu}\oplus{\cal H}_{-\Delta,-\delta\mu}$. The blocks can be expressed as ${\cal H}_{\Delta,\delta\mu}={\cal H}_\Delta-\delta\mu\ {\cal I}$ where ${\cal I}$ is the identity matrix and the matrix ${\cal H}_\Delta$ is given by
\begin{eqnarray}\label{highdensity}
({\cal H}_\Delta)_{{\bf G},{\bf G}^\prime}=\left(\begin{array}{cc} (|{\bf k}+{\bf G}|-\mu)\delta_{{\bf G},{\bf G}^\prime}
& \Delta_{{\bf G}-{\bf G}^\prime} \\ \Delta^*_{{\bf G}^\prime-{\bf G}}
& -(|{\bf k}+{\bf G}|-\mu)\delta_{{\bf G},{\bf G}^\prime}  \end{array}\right).
\end{eqnarray}
The eigenvalues of ${\cal H}_{\Delta,\delta\mu}$ do not depend on the sign of $\Delta$. Moreover,
replacing the $\delta\mu$ by $-\delta\mu$ amounts to a replacement of the eigenvalue spectrum $\{E_\lambda({\bf k})\}$ by
$\{-E_\lambda({\bf k})\}$. Therefore, the two blocks contribute equally to the grand potential and we only need to
determine the eigenvalues of ${\cal H}_{\Delta,\delta\mu}$. The Hamiltonian matrix (\ref{highdensity}) represents
the general problem of two-species pairing with mismatched Fermi surfaces. In the weak coupling limit, the pairing is dominated near
the Fermi surfaces. Therefore, the physical result should be universal in terms of the pairing gap $\Delta_0$ at vanishing mismatch
and the density of state ${\cal N}_{\rm F}$ at the Fermi surface.

\begin{figure}[!htb]
\begin{center}
\includegraphics[width=8.6cm]{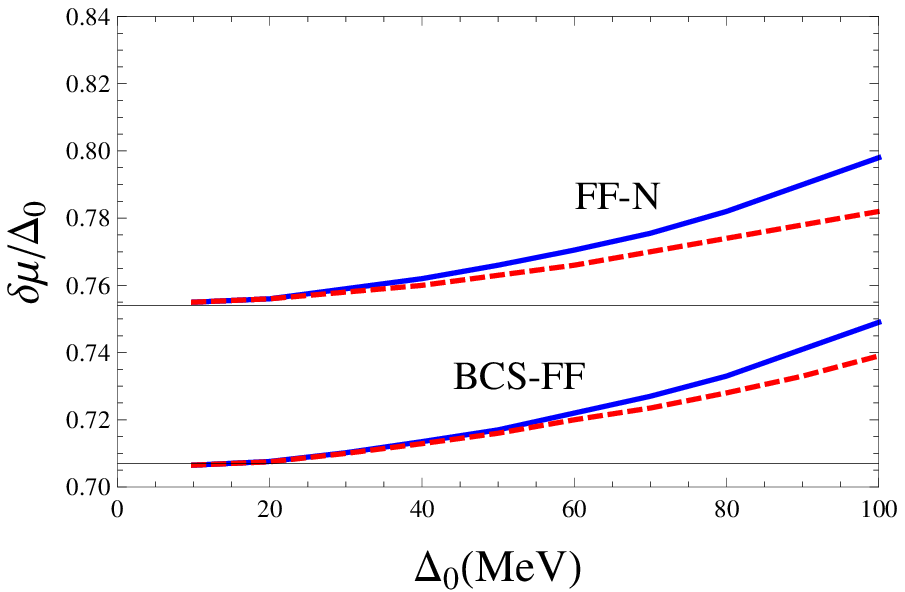}
\includegraphics[width=8.5cm]{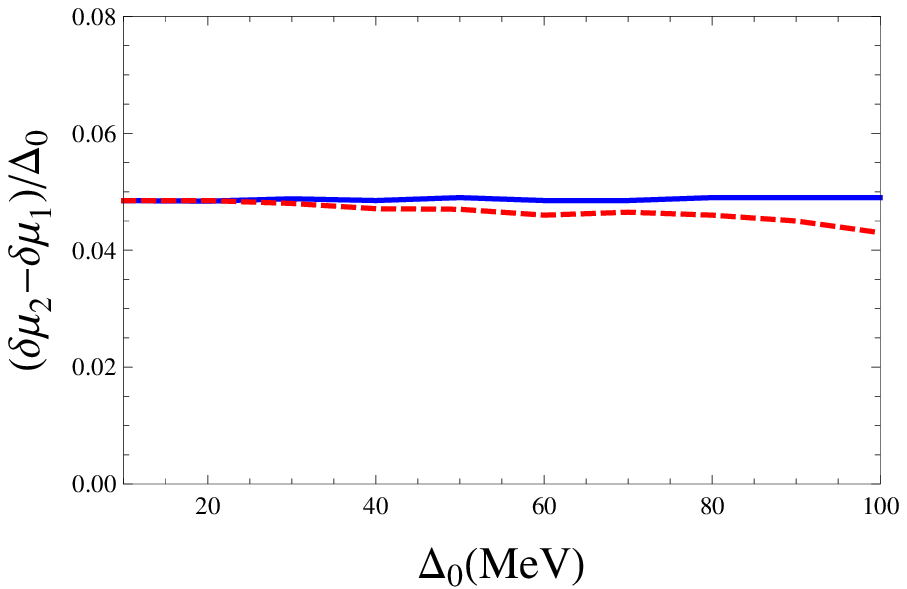}
\caption{(Color online) The lower and upper critical fields (upper panel) and the size of the stability window
$(\delta\mu_2-\delta\mu_1)/\Delta_0$ (lower panel) for the FF state as a function of $\Delta_0$ at $\mu=400$ MeV.
The thin lines denote results in the weak-coupling limit. The blue solid and red dashed lines correspond to
$\Lambda=400$ MeV and $\Lambda=800$ MeV, respectively. \label{fig1}}
\end{center}
\end{figure}

In the following we shall focus on the zero-temperature case. The grand potential $\Omega$ is divergent and hence a proper regularization scheme is needed. Since we need to deal with the Bloch momentum ${\bf k}+{\bf G}$, the usual three-momentum cutoff scheme~\cite{Alford2001,Bowers2002} is not appropriate for numerical calculations. Moreover, we are interested in the grand potential $\delta\Omega$ measured with respect to the normal state. Therefore, we employ a Pauli-Villars-like regularization scheme, in which $\delta\Omega$ is well-defined~\cite{Buballa2009}. The ``renormalized" grand potential is given by~\cite{note1}
\begin{eqnarray}
\delta\Omega(\Delta,q)=\Omega(\Delta,q)-\Omega(0,q),
\end{eqnarray}
where
\begin{eqnarray}
\Omega(\Delta,q)=\frac{P\Delta^2}{4g}-\frac{1}{2}\int_{\rm BZ}\frac{d^3{\bf k}}{(2\pi)^3}\sum_\lambda
\sum_{j=0}^2c_j\sqrt{E_\lambda^2({\bf k})+j\Lambda^2}
\end{eqnarray}
with $c_0=c_2=1$ and $c_1=-2$. Here $\{E_\lambda(\bf k)\}$ denotes the eigenvalue spectrum of ${\cal H}_{\Delta,\delta\mu}$. The coupling constant $g$ can be fixed by the BCS gap $\Delta_0$ at $\delta\mu=0$. We expect that at weak coupling the physical results depend on the cutoff $\Lambda$ only through the BCS gap $\Delta_0$~\cite{Buballa2009}. In Fig. \ref{fig1}, we show the stability window for the FF state as a function of $\Delta_0$. In the weak coupling limit, the critical fields depend only on $\Delta_0$. For accuracy reason~\cite{note2}, we shall choose $\Delta_0\sim100$ MeV at $\mu=400$ MeV, which corresponds to the realistic value of $\Delta_0$ at moderate density~\cite{CSC01}. Since the size of the FF window $(\delta\mu_2-\delta\mu_1)/\Delta_0$ depends very weakly on $\Delta_0$ and $\Lambda$, we can use the upper critical field $\delta\mu_2$ obtained in Fig. \ref{fig1} to ``calibrate" $\delta\mu$ and appropriately extrapolate the results to the weak coupling limit.

\section{Matrix Structure}

For a given ${\bf k}$-point in the BZ, we can diagonalize the Hamiltonian matrix ${\cal H}_{\Delta,\delta\mu}$ to obtain its eigenvalue spectrum $\{E_\lambda({\bf k})\}$. The choice of the ${\bf k}$-points in the BZ should be dense enough to achieve the thermodynamic limit~\cite{note3}.
The eigenvalue equation can be rewritten as
\begin{eqnarray}
\sum_{{\bf G}^\prime}({\cal H}_{\Delta})_{{\bf G},{\bf G}^\prime}({\bf k})\phi_{{\bf G}^\prime}({\bf k})
=\left[E_\lambda({\bf k})+\delta\mu\right] \phi_{\bf G}({\bf k}),
\end{eqnarray}
where the Hamiltonian matrix $({\cal H}_{\Delta})_{{\bf G},{\bf G}^\prime}({\bf k})$ reads
\begin{eqnarray}
({\cal H}_{\Delta})_{{\bf G},{\bf G}^\prime}({\bf k})=\left(\begin{array}{cc} \xi_{{\bf k}+{\bf G}}\delta_{{\bf G},{\bf G}^\prime} & \Delta_{{\bf G}-{\bf G}^\prime} \\ \Delta_{{\bf G}-{\bf G}^\prime}
& -\xi_{{\bf k}+{\bf G}}\delta_{{\bf G},{\bf G}^\prime}  \end{array}\right).
\end{eqnarray}
Here $\xi_{\bf p}=|{\bf p}|-\mu$ and we have used the fact $\Delta^*_{{\bf G}}=\Delta_{-{\bf G}}$.
The eigenstate $\phi_{\bf G}$ includes two components $u_{\bf G}$ and ${\bf \upsilon}_{\bf G}$ as usual in the BCS theory. We have
\begin{eqnarray}
\phi_{\bf G}({\bf k})=\left(\begin{array}{cc} u_{\bf G}({\bf k}) \\ \upsilon_{\bf G}({\bf k}) \end{array}\right).
\end{eqnarray}
We notice that $\delta\mu$ can be absorbed into the eigenvalues.  It is easy to prove that the eigenvalues of ${\cal H}_{\Delta}$ do not depend on the sign of $\Delta$. Moreover, if $\varepsilon$ is an eigenvalue of ${\cal H}_{\Delta}$, $-\varepsilon$ must be another eigenvalue. Therefore, replacing the $\delta\mu$ by $-\delta\mu$ amounts to a replacement of the eigenvalue spectrum $\{E_\lambda({\bf k})\}$ by
$\{-E_\lambda({\bf k})\}$.

However, the matrix ${\cal H}_\Delta$ has infinite dimensions because the integers $l,m,n$ run from $-\infty$ to $+\infty$. Therefore, we have to make a truncation in order to perform a calculation. It is natural to make a symmetrical truncation, i.e.,
\begin{equation}
-D\leq l,m,n\leq D,\ \ \ (D\in \mathbb{Z}^+).
\end{equation}
For sufficiently large $D$, the contribution from the high-energy bands becomes vanishingly small and the grand potential $\delta\Omega$ converges to its precise value. After making this truncation, the matrix equation can be expressed as
\begin{eqnarray}
{\bf H}\left(\begin{array}{cc} u \\ \upsilon \end{array}\right)=\left(\begin{array}{cc} {\bf H}_{11} & {\bf H}_{12} \\ {\bf H}_{21}
& {\bf H}_{22}  \end{array}\right)\left(\begin{array}{cc} u \\ \upsilon \end{array}\right)=(E+\delta\mu)\left(\begin{array}{cc} u \\ \upsilon \end{array}\right),
\end{eqnarray}
where $u$ and $\upsilon$ are $(2D+1)^3$-dimensional vectors and ${\bf H}_{ij}$ are $(2D+1)^3\times(2D+1)^3$ matrices. The matrix elements of
${\bf H}_{ij}$ can be formally expressed as
\begin{eqnarray}
&&{\bf H}_{11}^{[l,m,n],[l^\prime,m^\prime,n^\prime]}=-{\bf H}_{22}^{[l,m,n],[l^\prime,m^\prime,n^\prime]}
=\xi_{[l,m,n]}\delta_{l,l^\prime}\delta_{m,m^\prime}\delta_{n,n^\prime},\nonumber\\
&&{\bf H}_{12}^{[l,m,n],[l^\prime,m^\prime,n^\prime]}={\bf H}_{21}^{[l,m,n],[l^\prime,m^\prime,n^\prime]}
=\Delta_{[l-l^\prime,m-m^\prime,n-n^\prime]}, \label{Elements}
\end{eqnarray}
where
\begin{eqnarray}
\xi_{[l,m,n]}=\sqrt{\left(k_x+\frac{2\pi l}{a}\right)^2+\left(k_y+\frac{2\pi m}{a}\right)^2+\left(k_z+\frac{2\pi n}{a}\right)^2}-\mu.
\end{eqnarray}
Here the matrix index $[l,m,n]$ corresponds to the reciprocal lattice vector
${\bf G}_{[lmn]}=(2\pi/a)(l{\bf e}_x+m{\bf e}_y+n{\bf e}_z)$. It shows that the blocks ${\bf H}_{11}$ and ${\bf H}_{22}$ are diagonal.
The off-diagonal blocks ${\bf H}_{12}$ and ${\bf H}_{21}$ carry the information of the order parameter $\Delta$ and characterize the
crystal structure.

For a specific value of $D$, we can write down the explicit form of the vectors $u$ and $\upsilon$ and the matrices ${\bf H}_{ij}$. Here we use $D=1$ as an example. The vectors $u$ and $\upsilon$ are $27$-dimensional can be expressed as
\begin{eqnarray}\label{Basis}
u=\left(\begin{array}{cc} u_{[-1,-1]} \\ u_{[-1,0]} \\ u_{[-1,1]} \\ u_{[0,-1]}
\\ u_{[0,0]} \\ u_{[0,1]} \\ u_{[1,-1]} \\ u_{[1,0]} \\ u_{[1,1]}\end{array}\right),\ \ \ \ \ \ \ \ \ \ \ \ \ \ \ \ \ \
\upsilon=\left(\begin{array}{cc} \upsilon_{[-1,-1]} \\ \upsilon_{[-1,0]} \\ \upsilon_{[-1,1]} \\ \upsilon_{[0,-1]}
\\ \upsilon_{[0,0]} \\ \upsilon_{[0,1]} \\ \upsilon_{[1,-1]} \\ \upsilon_{[1,0]} \\ \upsilon_{[1,1]} \end{array}\right),
\end{eqnarray}
where $u_{[l,m]}$ and $\upsilon_{[l,m]}$ are defined as
\begin{eqnarray}
u_{[l,m]}=\left(\begin{array}{cc} u_{[l,m,-1]} \\ u_{[l,m,0]} \\ u_{[l,m,1]}\end{array}\right),\ \ \ \ \ \ \ \ \ \ \ \ \ \ \ \ \
\upsilon_{[l,m]}=\left(\begin{array}{cc} \upsilon_{[l,m,-1]} \\ \upsilon_{[l,m,0]} \\ \upsilon_{[l,m,1]} \end{array}\right).
\end{eqnarray}
In this representation, the off diagonal blocks ${\bf H}_{12}$ and ${\bf H}_{21}$ are given by
\begin{eqnarray}\label{MBCC}
{\bf H}_{12}=\left(\begin{array}{ccccccccc}
\mbox{\boldmath{$\Delta$}}_1  & \mbox{\boldmath{$\Delta$}}_2 & 0 & \mbox{\boldmath{$\Delta$}}_2 & 0 & 0 & 0 & 0 & 0 \\
\mbox{\boldmath{$\Delta$}}_2 & \mbox{\boldmath{$\Delta$}}_1 & \mbox{\boldmath{$\Delta$}}_2 & 0 & \mbox{\boldmath{$\Delta$}}_2 & 0 & 0 & 0 & 0\\
0 & \mbox{\boldmath{$\Delta$}}_2 & \mbox{\boldmath{$\Delta$}}_1 & 0 & 0 & \mbox{\boldmath{$\Delta$}}_2 & 0 & 0 & 0 \\
\mbox{\boldmath{$\Delta$}}_2 & 0 & 0 & \mbox{\boldmath{$\Delta$}}_1 & \mbox{\boldmath{$\Delta$}}_2 & 0 & \mbox{\boldmath{$\Delta$}}_2 & 0 & 0 \\
0 & \mbox{\boldmath{$\Delta$}}_2 & 0 & \mbox{\boldmath{$\Delta$}}_2 & \mbox{\boldmath{$\Delta$}}_1 & \mbox{\boldmath{$\Delta$}}_2 & 0 & \mbox{\boldmath{$\Delta$}}_2 & 0 \\
0 & 0 & \mbox{\boldmath{$\Delta$}}_2 & 0 & \mbox{\boldmath{$\Delta$}}_2 & \mbox{\boldmath{$\Delta$}}_1 & 0 & 0 & \mbox{\boldmath{$\Delta$}}_2  \\
0 & 0 & 0 & \mbox{\boldmath{$\Delta$}}_2 & 0 & 0 & \mbox{\boldmath{$\Delta$}}_1 & \mbox{\boldmath{$\Delta$}}_2 & 0  \\
0 & 0 & 0 & 0 & \mbox{\boldmath{$\Delta$}}_2 & 0 & \mbox{\boldmath{$\Delta$}}_2 & \mbox{\boldmath{$\Delta$}}_1 & \mbox{\boldmath{$\Delta$}}_2  \\
0 & 0 & 0 & 0 & 0 & \mbox{\boldmath{$\Delta$}}_2 & 0 & \mbox{\boldmath{$\Delta$}}_2 & \mbox{\boldmath{$\Delta$}}_1  \end{array}\right)
\end{eqnarray}
for BCC structure and
\begin{eqnarray}\label{MFCC}
{\bf H}_{12}=\left(\begin{array}{ccccccccc}
0 & 0 & 0 & 0 & \mbox{\boldmath{$\Delta$}}_1 & 0 & 0 & 0 & 0 \\
0 & 0 & 0 & \mbox{\boldmath{$\Delta$}}_1 & 0 & \mbox{\boldmath{$\Delta$}}_1 & 0 & 0 & 0\\
0 & 0 & 0 & 0 & \mbox{\boldmath{$\Delta$}}_1 & 0 & 0 & 0 & 0 \\
0 & \mbox{\boldmath{$\Delta$}}_1 & 0 & 0 & 0 & 0 & 0 & \mbox{\boldmath{$\Delta$}}_1 & 0 \\
\mbox{\boldmath{$\Delta$}}_1 & 0 & \mbox{\boldmath{$\Delta$}}_1 & 0 & 0 & 0 & \mbox{\boldmath{$\Delta$}}_1 & 0 & \mbox{\boldmath{$\Delta$}}_1 \\
0 & \mbox{\boldmath{$\Delta$}}_1 & 0 & 0 & 0 & 0 & 0 & \mbox{\boldmath{$\Delta$}}_1 & 0  \\
0 & 0 & 0 & 0 & \mbox{\boldmath{$\Delta$}}_1 & 0 & 0 & 0 & 0  \\
0 & 0 & 0 & \mbox{\boldmath{$\Delta$}}_1 & 0 & \mbox{\boldmath{$\Delta$}}_1 & 0 & 0 & 0  \\
0 & 0 & 0 & 0 & \mbox{\boldmath{$\Delta$}}_1 & 0 & 0 & 0 & 0  \end{array}\right)
\end{eqnarray}
for FCC structure, respectively. Here the blocks $\mbox{\boldmath{$\Delta$}}_1$ and $\mbox{\boldmath{$\Delta$}}_2$ are defined as
\begin{eqnarray}
\mbox{\boldmath{$\Delta$}}_1=\left(\begin{array}{ccc}
0 & \Delta & 0 \\ \Delta & 0 & \Delta \\ 0 & \Delta & 0\end{array}\right),\ \ \ \ \ \ \ \ \ \ \ \ \
\mbox{\boldmath{$\Delta$}}_2=\left(\begin{array}{ccc}
\Delta & 0 & 0 \\ 0 & \Delta & 0 \\ 0 & 0 & \Delta\end{array}\right).
\end{eqnarray}
In principle, the eigenvalue spectrum $\{E_\lambda({\bf k})\}$ can be obtained by diagonalizing the matrix ${\bf H}$ with a size $2(2D+1)^3$.

We notice that the matrix size $2(2D+1)^3$ grows dramatically with increasing cutoff $D$.  Therefore, for realistic diagonalization, it is better to reduce the size of the matrix. Here we find that, with a proper rearrangement of the basis $\phi$ or after a similarity transformation, the matrix ${\bf H}$ becomes block diagonal. We have
\begin{eqnarray}
{\bf H}\sim\left(\begin{array}{cc}
{\bf H}_+ & 0 \\ 0 & {\bf H}_- \end{array}\right),
\end{eqnarray}
where size of the blocks ${\bf H}_+$ and ${\bf H}_-$ are both $(2D+1)^3$. The eigenvector $\phi$ is now defined as
\begin{eqnarray}
\phi=\left(\begin{array}{cc}
\phi_+ \\ \phi_- \end{array}\right).
\end{eqnarray}
For $D=1$, the $27$-dimensional vectors $\phi_+$ and $\phi_-$ are given by
\begin{eqnarray}
\phi_+=\left(\begin{array}{cc} \upsilon_{[-1,-1,-1]} \\ u_{[-1,-1,0]} \\ \upsilon_{[-1,-1,1]} \\ u_{[-1,0,-1]} \\
\upsilon_{[-1,0,0]} \\ u_{[-1,0,1]} \\ \upsilon_{[-1,1,-1]} \\ u_{[-1,1,0]} \\ \upsilon_{[-1,1,1]} \\ u_{[0,-1,-1]} \\
\upsilon_{[0,-1,0]} \\ u_{[0,-1,1]} \\ \upsilon_{[0,0,-1]} \\ u_{[0,0,0]} \\ \upsilon_{[0,0,1]}\\ u_{[0,1,-1]} \\
\upsilon_{[0,1,0]} \\ u_{[0,1,1]} \\ \upsilon_{[1,-1,-1]} \\ u_{[1,-1,0]} \\ \upsilon_{[1,-1,1]} \\ u_{[1,0,-1]} \\
\upsilon_{[1,0,0]} \\ u_{[1,0,1]} \\ \upsilon_{[1,1,-1]} \\ u_{[1,1,0]} \\ \upsilon_{[1,1,1]}\end{array}\right),
\ \ \ \ \ \ \ \ \ \ \ \ \ \ \ \ \
\phi_-=\left(\begin{array}{cc} u_{[-1,-1,-1]} \\ \upsilon_{[-1,-1,0]} \\ u_{[-1,-1,1]} \\ \upsilon_{[-1,0,-1]} \\u_{[-1,0,0]} \\ \upsilon_{[-1,0,1]}
\\ u_{[-1,1,-1]} \\ \upsilon_{[-1,1,0]} \\ u_{[-1,1,1]} \\ \upsilon_{[0,-1,-1]} \\ u_{[0,-1,0]} \\ \upsilon_{[0,-1,1]}
\\ u_{[0,0,-1]} \\ \upsilon_{[0,0,0]} \\ u_{[0,0,1]}\\ \upsilon_{[0,1,-1]} \\ u_{[0,1,0]} \\ \upsilon_{[0,1,1]}
\\ u_{[1,-1,-1]} \\ \upsilon_{[1,-1,0]} \\ u_{[1,-1,1]} \\ \upsilon_{[1,0,-1]} \\ u_{[1,0,0]} \\ \upsilon_{[1,0,1]}
\\ u_{[1,1,-1]} \\ \upsilon_{[1,1,0]} \\ u_{[1,1,1]}\end{array}\right),
\end{eqnarray}
which is just a proper rearrangement of the original basis given by (\ref{Basis}). The blocks ${\bf H}_+$ and ${\bf H}_-$ are given by
\begin{eqnarray}
{\bf H}_\pm=\pm{\bf H}_0+{\bf H}_{12},
\end{eqnarray}
where ${\bf H}_{12}$ is given by (\ref{Elements}) or (\ref{MBCC}) and (\ref{MFCC}) for $D=1$. ${\bf H}_0$ is a diagonal matrix containing the kinetic energies $\xi_{[l,m,n]}$. We have
\begin{eqnarray}
{\bf H}_0^{[l,m,n],[l^\prime,m^\prime,n^\prime]}
=(-1)^{l+m+n}\xi_{[l,m,n]}\delta_{l,l^\prime}\delta_{m,m^\prime}\delta_{n,n^\prime}.
\end{eqnarray}
For $D=1$, we obtain
\begin{eqnarray}
{\bf H}_0&=&{\rm diag}(-\xi_{[-1,-1,-1]},\xi_{[-1,-1,0]},-\xi_{[-1,-1,1]},\cdots,\nonumber\\
&&\ \ \ \ \ \ \ \ \xi_{[0,0,0]},\cdots,-\xi_{[1,1,-1]},\xi_{[1,1,0]},-\xi_{[1,1,1]}).
\end{eqnarray}
\begin{figure*}[!htb]
\begin{center}
\includegraphics[width=7.8cm]{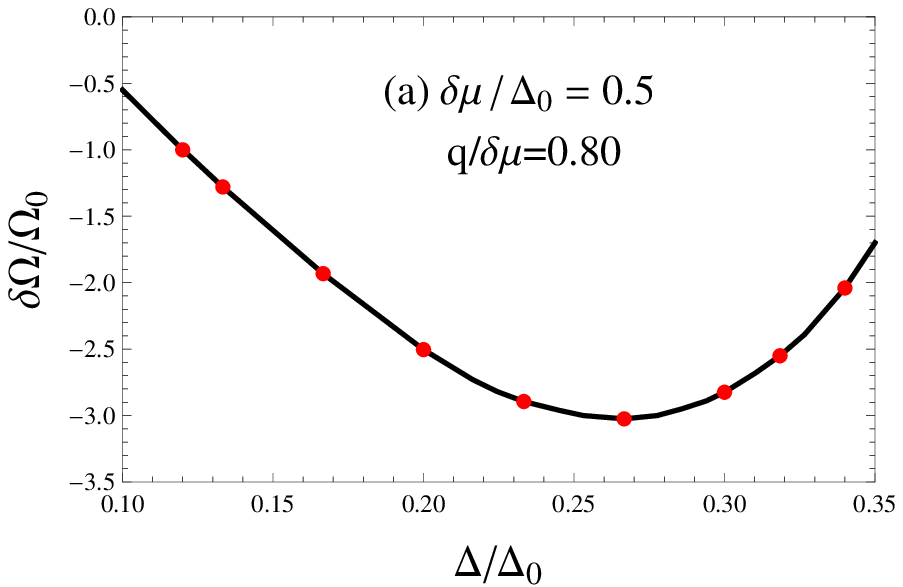}
\includegraphics[width=7.8cm]{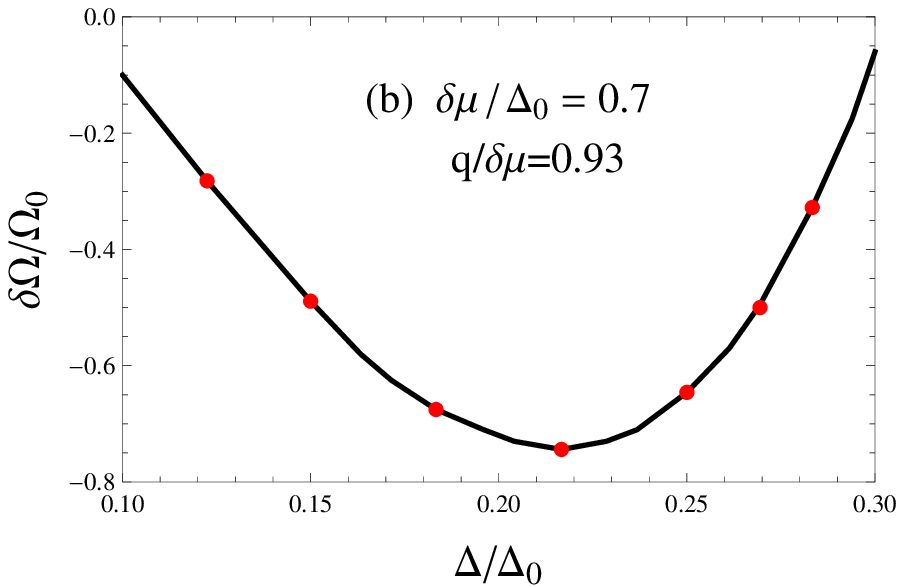}
\includegraphics[width=7.8cm]{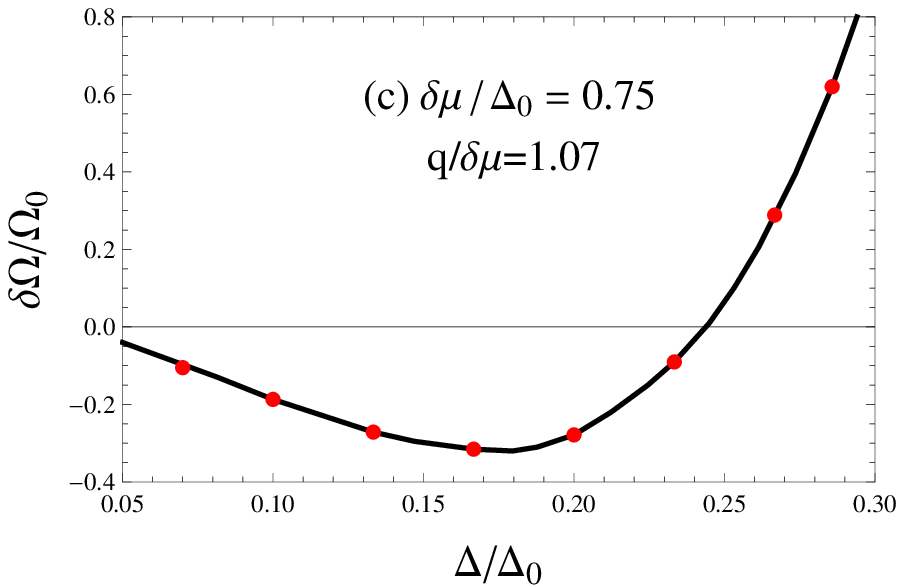}
\includegraphics[width=7.8cm]{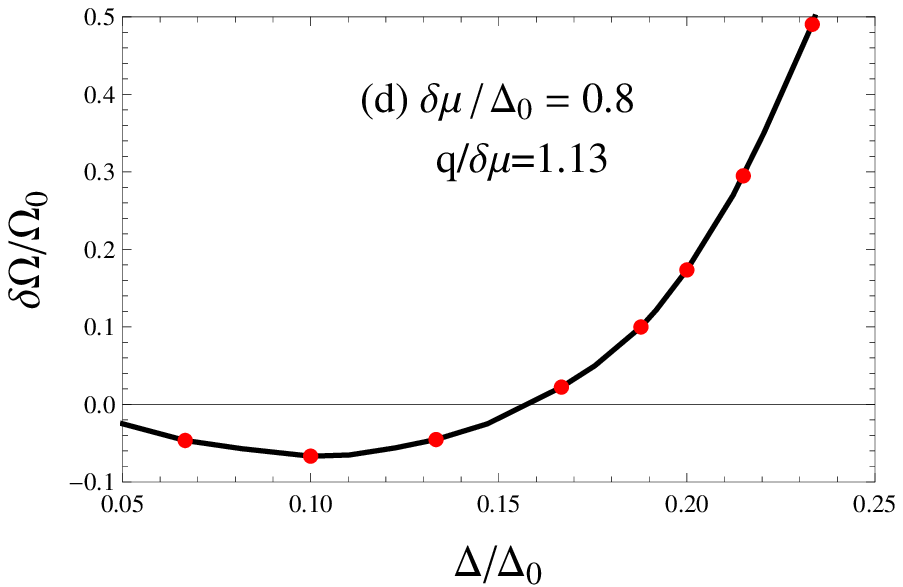}
\caption{(Color online) The potential curves $\delta\Omega(\Delta)$ of the BCC structure at the optimal pair momenta for various values of $\delta\mu/\Delta_0$. The grand potential is scaled by a constant
$\Omega_0=2.5\times10^6 ({\rm MeV})^4$. The red dots show the data obtained from our numerical calculation. \label{fig2}}
\end{center}
\end{figure*}

It is easy to show that the eigenvalue spectra of ${\bf H}_+$ and ${\bf H}_-$ are dependent: If the eigenvalue spectrum of ${\bf H}_+$ is given by
$\{\varepsilon_\lambda({\bf k})\}$, the eigenvalue spectrum of ${\bf H}_-$ reads $\{-\varepsilon_\lambda({\bf k})\}$. Therefore, we only need to diagonalize the matrix ${\bf H}_+$ or ${\bf H}_-$ which has a size $(2D+1)^3$. Once the eigenvalue spectrum of ${\bf H}_+$ is known, the eigenvalue spectrum of Hamiltonian matrix ${\cal H}_{\Delta,\delta\mu}$ is given by
\begin{eqnarray}
\{E_\lambda({\bf k})\}=\{\varepsilon_\lambda({\bf k})-\delta\mu\}\cup\{-\varepsilon_\lambda({\bf k})-\delta\mu\}.
\end{eqnarray}
Therefore, we can in principle calculate the potential landscape $\delta\Omega(\Delta,q)$. The solution $(\Delta,q)$ of a specific crystal structure corresponds to the global minimum of the potential landscape.

\section{Computation and Results}

To achieve satisfying convergence we normally need a large cutoff $D$. However, the matrix size $(2D+1)^3$ and hence the computing time and cost grow dramatically with increasing $D$.  The cutoff $D$ needed for convergence can be roughly estimated from the maximum momentum $k_{\rm max}$ in the matrix,
\begin{equation}
k_{\rm max}=(2D+1)\frac{\pi}{a}
\end{equation}
The value of $k_{\rm max}$ can be estimated from the LO state. The calculation of the LO state is much easier than 3D structures because the matrix size becomes $2D+1$. The details of the calculation of the LO state are presented in Appendix B. For $\Delta_0\sim100$ MeV we need $k_{\rm max}\simeq5$GeV~\cite{note4}. Since we are interested in the region $\delta\mu/\Delta_0\in[0.7,0.8]$ and the optimal pair momentum is $q\sim\delta\mu$, we estimate $D\sim35$ for BCC and $D\sim60$ for FCC. These huge matrix sizes are beyond the capability of our current computing facilities. On the other hand, even though a supercomputer may be able to diagonalize these huge matrices, the computing time and cost are still enormous, which makes the calculation infeasible.

\begin{figure}[!htb]
\begin{center}
\includegraphics[width=8.7cm]{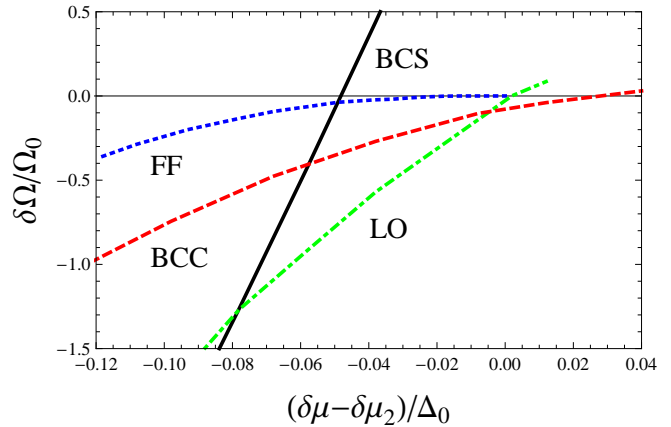}
\caption{(Color online) Comparison of the grand potentials of various phases: BCS (black solid), FF (blue dotted),
LO (green dash-dotted), and BCC (red dashed). The horizontal axis has been ``calibrated" by using the quantity
$(\delta\mu-\delta\mu_2)/\Delta_0$.
\label{fig3}}
\end{center}
\end{figure}

Since we are interested in the grand potential $\delta\Omega$ rather than the band structure (the eigenvalues), we can neglect a small amount of the off-diagonal couplings $\Delta$ in the matrix ${\bf H}_+$. By doing so, the huge matrix ${\bf H}_+$ can be decomposed into a number of blocks with size $(2d+1)^3$. For symmetry reason, we set the centers of these blocks at the reciprocal lattice vectors
\begin{equation}
{\bf G}_{[n_x,n_y,n_z]}=(2d+1)\frac{2\pi}{a}(n_x{\bf e}_x+n_y{\bf e}_y+n_z{\bf e}_z)
\end{equation}
with $n_x,n_y,n_z\in\mathbb{Z}$. With increasing $d$, the grand potential converges to the result from exact diagonalization. Good convergence is normally reached at some value $d=d_0$. The details of our computational scheme are presented in Appendix C. If the block size $(2d_0+1)^3$ is within our computing capability, the calculation becomes feasible. Fortunately, we find that this computational scheme works for the BCC structure. At present, we are not able to perform a calculation for the FCC structure, since the value of $d_0$ needed for convergence
is much larger. Note that the computing cost is still very large even though we have employed this effective computational scheme.

We have performed calculations of the BCC structure for $\Delta_0=60,80,100$ MeV~\cite{note2} at $\mu=400$ MeV~\cite{note5}. For different values of $\Delta_0$, the results are almost the same in terms of the quantity $(\delta\mu-\delta\mu_2)/\Delta_0$. Therefore, we anticipate that our results can be appropriately extrapolated to the weak coupling limit $\Delta_0\rightarrow0$. In the following, we shall present the result for $\Delta_0=100$ MeV. For a given value of $\delta\mu/\Delta_0$, we calculate the potential curve $\delta\Omega(\Delta)$ at various values of $q$ and search for the optimal pair momentum and the minimum of the potential landscape. The potential curves at the optimal pair momenta for several values of $\delta\mu/\Delta_0$ are shown in Fig. \ref{fig2}. With increasing value of $\delta\mu/\Delta_0$, the potential minimum gets shallower. At a critical value $\delta\mu_*-\delta\mu_2\simeq0.03\Delta_0$, the potential minimum approaches zero and a first-order phase transition to the normal state occurs. The comparison of the grand potentials of various phases are shown in Fig. \ref{fig3}. For the LO state, its phase transition to the normal state occur almost at the same point as the FF state, $\delta\mu_2\simeq0.8\Delta_0$.
At $\delta\mu=\delta\mu_2$, the grand potential of the BCC structure is negative, which indicates that the BCC structure is energetically favored around the FF-normal transition point. Well below the FF-normal transition point, the BCC state has higher grand potential than the LO state and hence is not favored. Near the BCS-LO transition, the solitionic state becomes favored~\cite{Buballa2009}. However, this does not change our qualitative conclusion.

Our result is qualitatively consistent with the GL analysis~\cite{Bowers2002}. However, the quantitative difference is significant: The GL analysis predicts a strong first-order phase transition and a large upper critical field~\cite{Bowers2002}, while our result shows a weak first-order phase transition at which $\Delta\simeq0.1\Delta_0$. On the other hand, our result is quantitatively compatible with the quasiclassical equation approach~\cite{Combescot2004,Combescot2005}, where it shows that the BCC structure is preferred in a narrow window around
$\delta\mu=\delta\mu_2$ at zero temperature~\cite{note6}. Therefore, the GL analysis up to the order $O(\Delta^6)$ may not be quantitatively sufficient. We notice that the LO state already shows the limitation of the GL analysis: While the GL analysis predicts a second-order phase transition, exact calculation shows a first-order phase transition~\cite{Buballa2009} (see also Appendix B). In the future, it is necessary to study the higher-order expansions and the convergence property of the GL series, which would help to quantitatively improve the GL predictions.

\section{Summary}

In summary, we have performed an solid-state-like calculation of the ground-state energy of a 3D structure in crystalline color superconductivity . We proposed a computational scheme to overcome the difficulties in diagonalizing matrices of huge sizes. Our numerical results show that the BCC structure is preferred in a small window around the conventional FF-normal phase transition point, which indicates that the higher-order terms in the GL approach are rather important. In the future it would be possible to perform a calculation for the FCC structure with stronger computing facilities and/or with better method of matrix diagonalization. This solid-state-like approach can also be applied to study the crystalline structures of the three-flavor color-superconducting quark matter~\cite{Three-flavor} and the inhomogeneous chiral condensate~\cite{Buballa2014}.

\emph{Acknowledgments} --- We thank Profs. Mark Alford, Joseph Carlson, Roberto Casalbuoni, Stefano Gandolfi, Hui Hu, Xu-Guang Huang, Massimo Mannarelli, Sanjay Reddy, Armen Sedrakian, and Shiwei Zhang for useful discussions and comments. The work of G. C. and P. Z. was supported by the NSFC under Grant No. 11335005 and the MOST under Grant Nos. 2013CB922000 and 2014CB845400. The work of L. H. was supported by the US Department
of Energy Topical Collaboration ``Neutrinos and Nucleosynthesis in Hot and Dense Matter". L. H. also acknowledges the support from Frankfurt Institute for Advanced Studies in the early stage of this work. The numerical calculations were performed at Tsinghua National Laboratory for Information Science and Technology.

\vspace{0.05in}
\appendix

\section{Ginzburg-Landau Theory: Importance of Higher Order Expansions}

In the Ginzburg-Landau (GL) theory of crystalline color superconductors at zero temperature and at weak coupling, the grand potential measured with respect to the normal state, $\delta\Omega=\Omega-\Omega_{\rm N}$, is expanded as~\cite{Bowers2002}
\begin{eqnarray}
\frac{\delta\Omega(\Delta)}{{\cal N}_{\rm F}}=P\alpha\Delta^2+\frac{1}{2}\beta\Delta^4
+\frac{1}{3}\gamma\Delta^6+\frac{1}{4}\eta\Delta^8+O(\Delta^{10}),
\end{eqnarray}
where ${\cal N}_{\rm F}$ is the density of state at the Fermi surface. The coefficient $\alpha$ is universal for all crystal structures and is given by~\cite{Bowers2002}
\begin{eqnarray}
\alpha=-1+\frac{\delta\mu}{2q}\ln\frac{q+\delta\mu}{q-\delta\mu}-\frac{1}{2}\ln\frac{\Delta_0^2}{4(q^2-\delta\mu^2)}.
\end{eqnarray}
\begin{figure*}[!htb]
\begin{center}
\includegraphics[width=7.8cm]{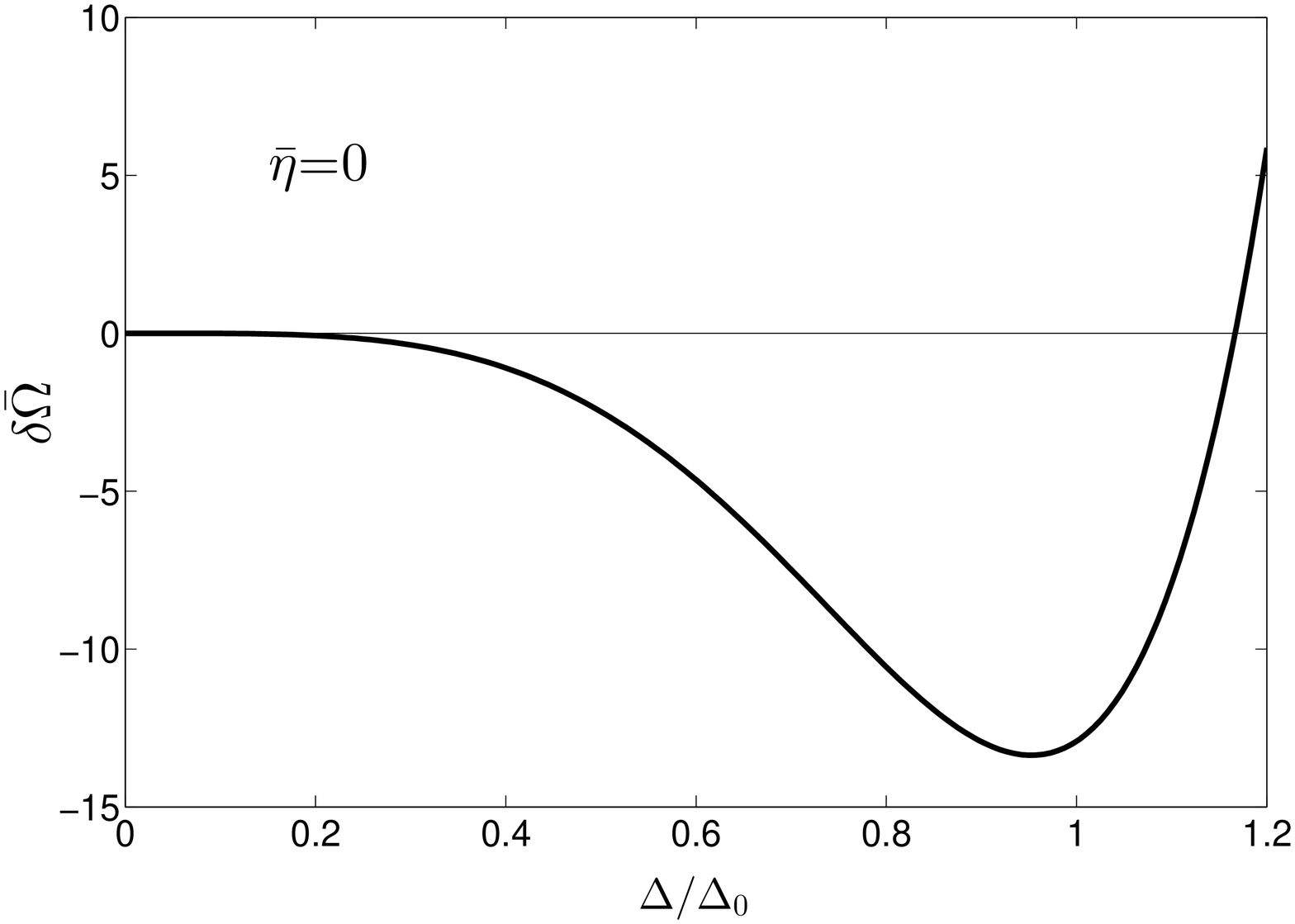}
\includegraphics[width=7.8cm]{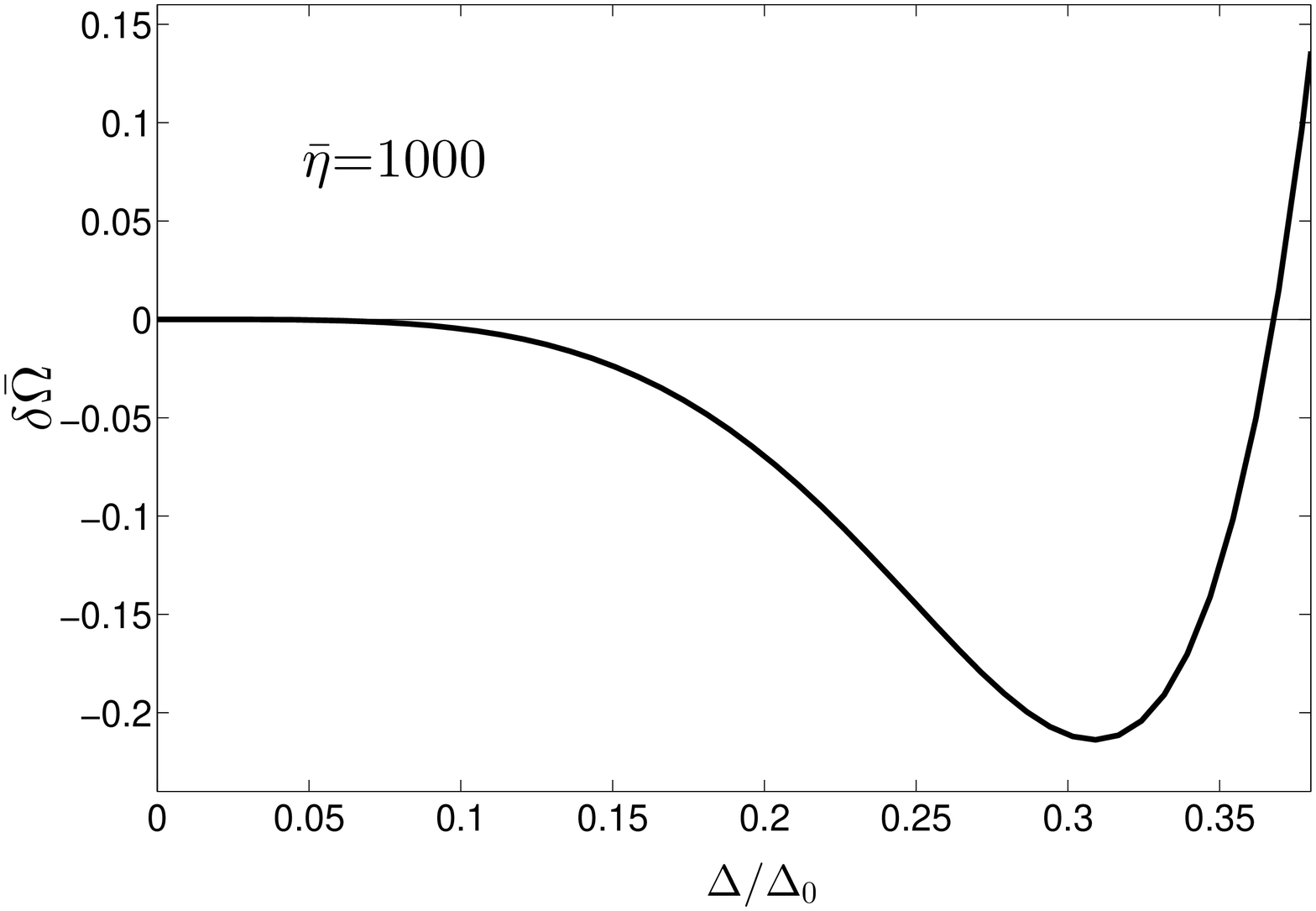}
\caption{The GL potential curves of the BCC structure for different values of $\bar{\eta}$ at $\delta\mu=\delta\mu_2=0.754\Delta_0$. \label{GLBCC}}
\end{center}
\end{figure*}
\begin{figure*}[!htb]
\begin{center}
\includegraphics[width=7.8cm]{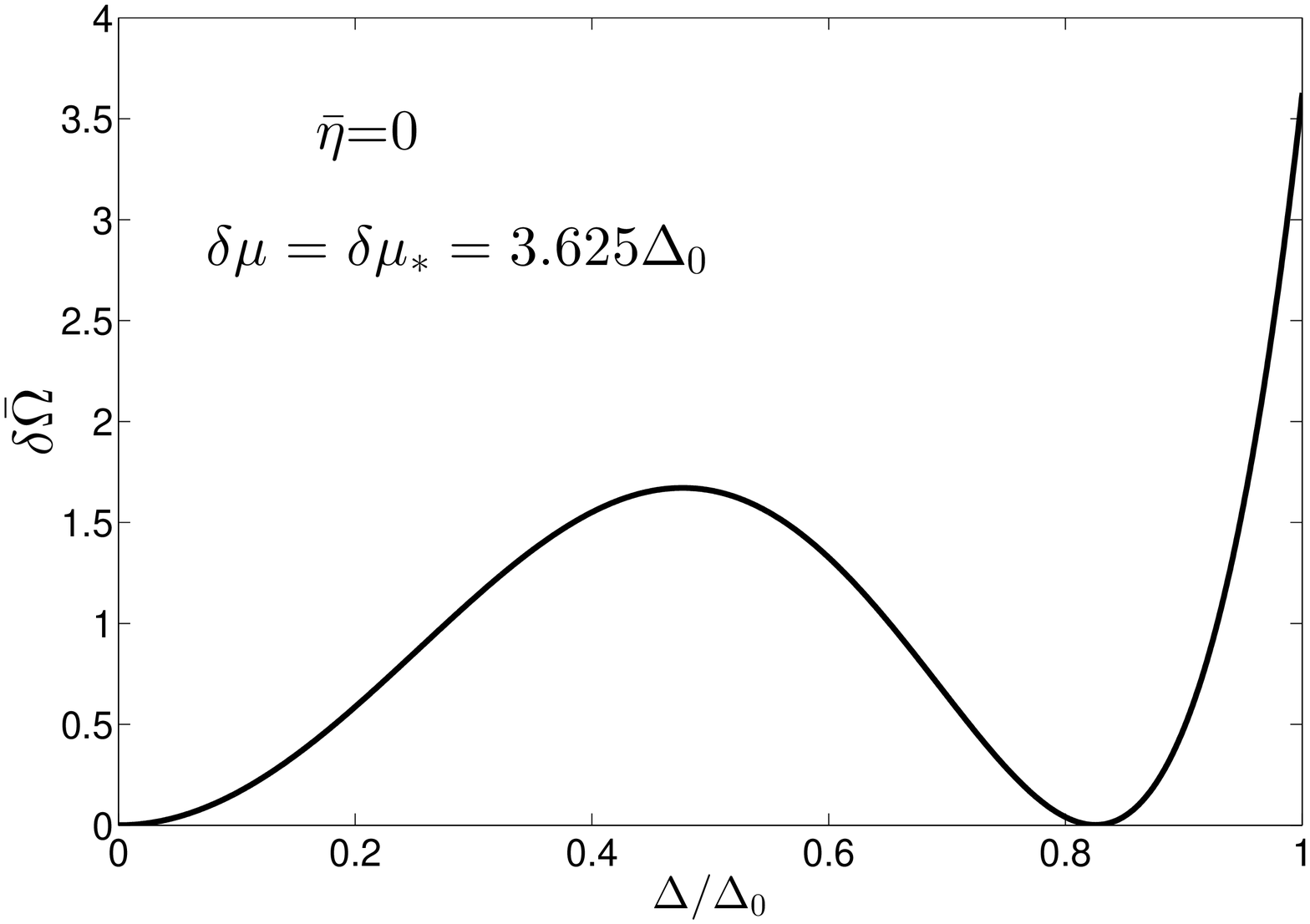}
\includegraphics[width=7.8cm]{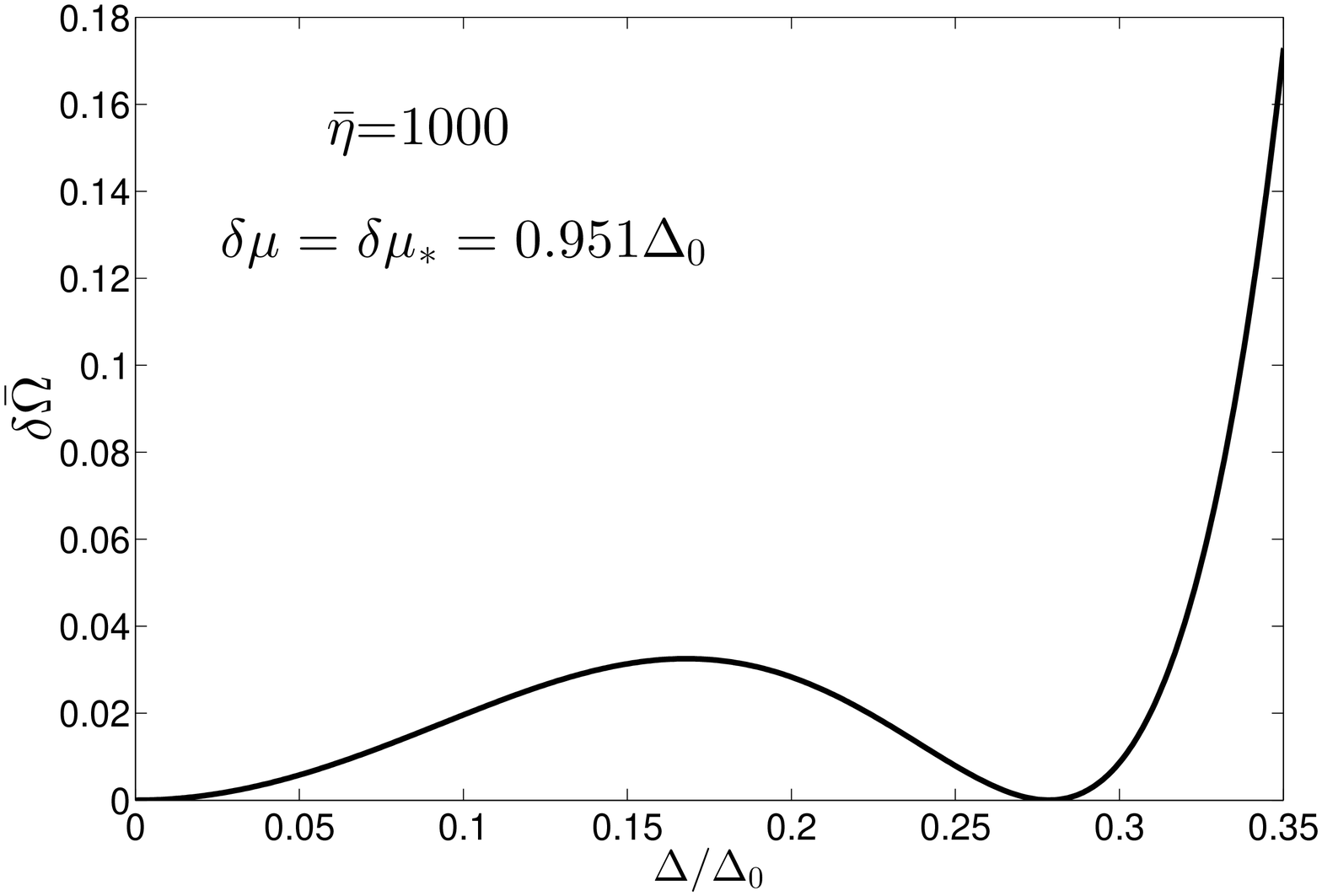}
\caption{The GL potential curves of the BCC structure for different values of $\bar{\eta}$ at the first-order phase transition point $\delta\mu=\delta\mu_*$. \label{GLBCC2}}
\end{center}
\end{figure*}
\\ Let us consider the vicinity of the conventional LOFF-normal transition point $\delta\mu=\delta\mu_2$, where we have
\begin{eqnarray}
\frac{\delta\mu_2}{\Delta_0}=0.7544,\ \ \ \ \ \frac{q}{\delta\mu_2}=1.1997.
\end{eqnarray}
At the pair momentum $q=1.1997\delta\mu$, we obtain
\begin{eqnarray}
\alpha=\ln\frac{\delta\mu}{\Delta_0}-\ln\frac{\delta\mu_2}{\Delta_0}=\ln\frac{\delta\mu}{\delta\mu_2}.
\end{eqnarray}
For convenience, we make the GL potential dimensionless by using the variables $\delta\bar{\Omega}=\delta\Omega/(N_0\delta\mu^2)$, $\bar{\Delta}=\Delta/\delta\mu$, $\bar{\beta}=\beta\delta\mu^2$, $\bar{\gamma}=\gamma\delta\mu^4$, and $\bar{\eta}=\eta\delta\mu^6$. We have
\begin{eqnarray}
\delta\bar{\Omega}=P\alpha\bar{\Delta}^2+\frac{1}{2}\bar{\beta}\bar{\Delta}^4+\frac{1}{3}\bar{\gamma}\bar{\Delta}^6
+\frac{1}{4}\bar{\eta}\bar{\Delta}^8+O(\bar{\Delta}^{10}).
\end{eqnarray}

The GL coefficients $\bar{\beta}$ and $\bar{\gamma}$ for a number of crystalline structures were first calculated by Bowers and Rajagopal~\cite{Bowers2002}. The predictions for the nature of the phase transitions were normally based on the GL potential up to the sixth order ($\Delta^6$). To our knowledge, the higher order GL coefficients have never been calculated. Here we show that the higher-order GL expansions are important for the prediction of the phase transition. To be specific, let us consider the BCC structure. Its GL coefficients $\bar{\beta}$ and $\bar{\gamma}$ have been evaluated as~\cite{Bowers2002}
\begin{eqnarray}
\bar{\beta}=-31.466,\ \ \ \ \ \bar{\gamma}=19.711.
\end{eqnarray}
Since $\bar{\beta}<0$, the phase transition to the normal state should be of first order. If we employ the GL potential up to the sixth order, we predict a strong first-order phase transition at $\delta\mu=\delta\mu_*=3.625\Delta_0$. Let us turn on the eighth-order term and study how the size of the coefficient $\bar{\eta}$ influences the quantitative prediction of the phase transition. In Fig. \ref{GLBCC}, we show the GL potential curves for two different values of $\bar{\eta}$ at $\delta\mu=\delta\mu_2$. For vanishing $\bar{\eta}$, the potential curve develops a deep minimum
$\delta\bar{\Omega}_{\rm min}\simeq-13.4$ at $\Delta\simeq0.95\Delta_0$, which indicates a strong first-order phase transition at $\delta\mu=\delta\mu_*\gg\delta\mu_2$. However, for a large value $\bar{\eta}=1000$, we find a shallow minimum $\delta\bar{\Omega}_{\rm min}\simeq-0.21$ located at $\Delta\simeq0.31\Delta_0$. In Fig. \ref{GLBCC2}, we show the GL potential curves at the first-order phase transition point $\delta\mu=\delta\mu_*$. For $\bar{\eta}=0$ we find a strong
first-order phase transition at $\delta\mu=\delta\mu_*=3.625\Delta_0$, where the minima located at $\Delta=0$ and $\Delta=0.83\Delta_0$ become degenerate. For $\bar{\eta}=1000$, however, we observe a much weaker first-order phase transition at $\delta\mu=\delta\mu_*=0.951\Delta_0$, where the degenerate minima are located at $\Delta=0$ and $\Delta=0.28\Delta_0$. These results clearly show that, for larger $\bar{\eta}$, the first-order phase transition becomes weaker.  For $\bar{\eta}\rightarrow+\infty$, we expect that $\delta\mu_*\rightarrow\delta\mu_2=0.754\Delta_0$. On the other hand, if $\bar{\eta}$ is small or even negative, then the next order $\Delta^{10}$ would become important.

\section{Calculation of the LO State}

The order parameter of the LO state is given by
\begin{eqnarray}
\Delta(z)=2\Delta\cos(2qz).
\end{eqnarray}
It is periodic along the $z$ direction with the periodicity $a=\pi/q$. So it can be decomposed into a discrete set of Fourier components,
\begin{eqnarray}
\Delta(z)=\sum_{n=-\infty}^\infty\Delta_ne^{2niqz},
\end{eqnarray}
The Fourier component $\Delta_n$ is given by
\begin{eqnarray}
\Delta_n=\frac{1}{a}\int_0^a dz\Delta(z)e^{-2niqz}=\Delta\left(\delta_{n,1}+\delta_{n,-1}\right).
\end{eqnarray}
The matrix equation takes a similar form as the 3D structures. We have
\begin{eqnarray}
\sum_{n^\prime}({\cal H}_{\Delta})_{n,n^\prime}({\bf k})\phi_{n^\prime}({\bf k})
=\left[E_\lambda({\bf k})+\delta\mu\right] \phi_n({\bf k}),
\end{eqnarray}
where the Hamiltonian matrix $({\cal H}_{\Delta})_{n,n^\prime}({\bf k})$ reads
\begin{eqnarray}
({\cal H}_{\Delta})_{n,n^\prime}({\bf k})=\left(\begin{array}{cc} \xi_n\delta_{n,n^\prime} & \Delta_{n-n^\prime} \\ \Delta_{n-n^\prime}
& -\xi_n\delta_{n,n^\prime}  \end{array}\right)
\end{eqnarray}
with
\begin{eqnarray}
\xi_n=\sqrt{{\bf k}_\perp^2+(k_z+2nq)^2}-\mu.
\end{eqnarray}
We notice that only the motion in the $z$ direction becomes quantized. The BZ for $k_z$ can be defined as $-\pi/a<k_z<\pi/a$ or $-q<k_z<q$.
The eigenstate $\phi_n$ includes two components $u_n$ and ${\bf \upsilon}_n$. We have
\begin{eqnarray}
\phi_n({\bf k})=\left(\begin{array}{cc} u_n({\bf k}) \\ \upsilon_n({\bf k}) \end{array}\right).
\end{eqnarray}
If $\varepsilon$ is an eigenvalue of ${\cal H}_{\Delta}$, $-\varepsilon$ must be another eigenvalue. Therefore, replacing the $\delta\mu$ by $-\delta\mu$ amounts to a replacement of the eigenvalue spectrum $\{E_\lambda({\bf k})\}$ by $\{-E_\lambda({\bf k})\}$.

After a truncation $-D<n<D$, we obtain a finite matrix equation
\begin{eqnarray}
{\bf H}\left(\begin{array}{cc} u \\ \upsilon \end{array}\right)=\left(\begin{array}{cc} {\bf H}_{11} & {\bf H}_{12} \\ {\bf H}_{21}
& {\bf H}_{22}  \end{array}\right)\left(\begin{array}{cc} u \\ \upsilon \end{array}\right)
=(E+\delta\mu)\left(\begin{array}{cc} u \\ \upsilon \end{array}\right),
\end{eqnarray}
where $u$ and $\upsilon$ are $(2D+1)$-dimensional vectors and ${\bf H}_{ij}$ are $(2D+1)\times(2D+1)$ matrices. For a specific value of $D$, we can write down the explicit form of the vectors $u$ and $\upsilon$ and the matrices ${\bf H}_{ij}$. Here we use $D=2$ as an example. The vectors $u$ and $\upsilon$ are $5$-dimensional can be expressed as
\begin{eqnarray}
u=\left(\begin{array}{cc} u_{-2} \\ u_{-1} \\ u_{0} \\ u_{1} \\ u_{2}\end{array}\right),
\ \ \ \ \ \ \ \ \ \ \ \ \ \ \ \ \ \ \ \ \
\upsilon=\left(\begin{array}{cc} \upsilon_{-2} \\ \upsilon_{-1} \\ \upsilon_{0} \\ \upsilon_{1} \\ \upsilon_2 \end{array}\right).
\end{eqnarray}
The matrices ${\bf H}_{ij}$ can be explicitly written as
\begin{eqnarray}
&&{\bf H}_{11}=-{\bf H}_{22}=\left(\begin{array}{ccccc} \xi_{-2} & 0 & 0 & 0 & 0 \\
0 & \xi_{-1} & 0 & 0 & 0 \\ 0 & 0 & \xi_0 & 0 & 0 \\ 0 & 0 & 0 & \xi_1 & 0 \\ 0 & 0 & 0 & 0 & \xi_2\end{array}\right), \nonumber\\
&&{\bf H}_{12}={\bf H}_{21}=\left(\begin{array}{ccccc} 0 & \Delta & 0 & 0 & 0 \\
\Delta & 0 & \Delta & 0 & 0 \\ 0 & \Delta & 0 & \Delta & 0 \\ 0 & 0 & \Delta & 0 & \Delta \\ 0 & 0 & 0 & \Delta & 0
 \end{array}\right).
\end{eqnarray}
The eigenvalue spectrum $\{E_\lambda({\bf k})\}$ can be obtained by diagonalizing the matrix ${\bf H}$ with a size $2(2D+1)$. With a proper rearrangement of the basis $\phi$ or a similarity transformation, we have
\begin{eqnarray}
{\bf H}\sim\left(\begin{array}{cc}
{\bf H}_+ & 0 \\ 0 & {\bf H}_- \end{array}\right),
\end{eqnarray}
where the sizes of ${\bf H}_+$ and ${\bf H}_-$ are both $2D+1$. The basis $\phi$ is now defined as
\begin{eqnarray}
\phi=\left(\begin{array}{cc}
\phi_+ \\ \phi_- \end{array}\right).
\end{eqnarray}
For $D=2$, the $5$-dimensional vectors $\phi_+$ and $\phi_-$ are given by
\begin{eqnarray}
\phi_+=\left(\begin{array}{cc} u_{-2} \\ \upsilon_{-1} \\ u_{0} \\ \upsilon_{1} \\ u_{2}\end{array}\right),
\ \ \ \ \ \ \ \ \ \ \ \ \ \ \ \
\phi_-=\left(\begin{array}{cc} \upsilon_{-2} \\ u_{-1} \\ \upsilon_{0} \\ u_{1} \\ \upsilon_2 \end{array}\right).
\end{eqnarray}
The blocks ${\bf H}_+$ and ${\bf H}_-$ can be expressed as
\begin{eqnarray}
{\bf H}_\pm=\pm{\bf H}_0+{\bf H}_{12}.
\end{eqnarray}
${\bf H}_0$ is a diagonal matrix containing the kinetic energies. We have
\begin{eqnarray}
({\bf H}_0)_{n,n^\prime}=(-1)^{n}\xi_{n}\delta_{n,n^\prime}.
\end{eqnarray}
The eigenvalue spectra of ${\bf H}_+$ and ${\bf H}_-$ are dependent: If the eigenvalue spectrum of ${\bf H}_+$ is given by $\{\varepsilon_\lambda({\bf k})\}$, the eigenvalue spectrum of ${\bf H}_-$ reads $\{-\varepsilon_\lambda({\bf k})\}$. Therefore, we need only to diagonalize the matrix ${\bf H}_+$ or ${\bf H}_-$ which has a dimension $2D+1$. Once the eigenvalue spectrum of ${\bf H}_+$ is known, the eigenvalue spectrum of Hamiltonian matrix ${\cal H}_{\Delta,\delta\mu}$ is given by
$\{E({\bf k})\}=\{\varepsilon_\lambda({\bf k})-\delta\mu\}\cup\{-\varepsilon_\lambda({\bf k})-\delta\mu\}$.

\begin{figure*}[!htb]
\begin{center}
\includegraphics[width=8cm]{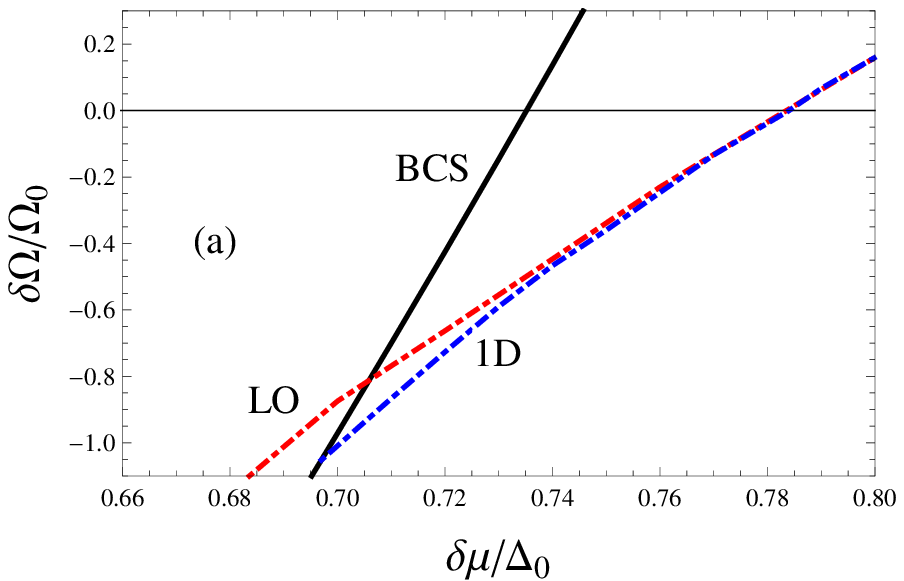}
\includegraphics[width=8cm]{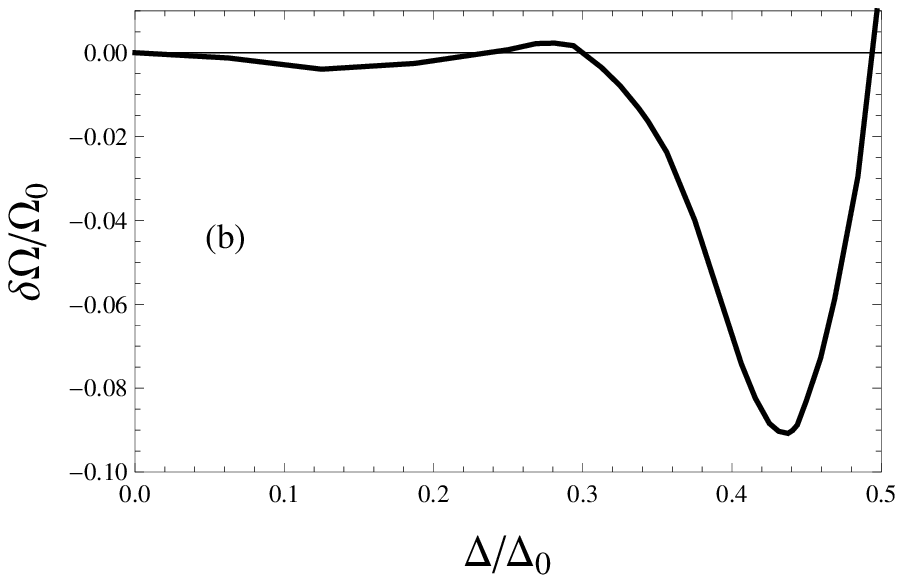}
\caption{(Color Online) (a) Comparison of the grand potentials of the LO state and the self-consistent 1D modulation for $\Delta_0=80$ MeV. (b) The potential curve of the LO state at $\delta\mu=0.775\Delta_0$ and at the optimal pair momentum $q=0.9\Delta_0$ or $q=1.1613\delta\mu$.  \label{LO}}
\end{center}
\end{figure*}

The thermodynamic potential of the LO state at zero temperature can be expressed as
\begin{eqnarray}
\Omega_{\rm LO}=\frac{\Delta^2}{2H}-2\int\frac{d^2{\bf k}_\perp}{(2\pi)^2}\int_{\rm BZ}\frac{dk_z}{2\pi}\sum_\lambda|E_\lambda({\bf k}_\perp, k_z)|.
\end{eqnarray}
Similar Pauli-Villas-like regularization scheme should be applied finally. In Fig. \ref{LO} (a), we show the grand potential of the LO state for $\Delta_0=80$ MeV. The grand potential for the self-consistent 1D modulation for $\Delta_0=80$ MeV was also reported in \cite{Buballa2009}. We find that the results for the LO state and the self-consistent 1D modulation agrees with each other near the phase transition to the normal state. Near the BCS-LO transition point, the self-consistent 1D modulation has lower grand potential than the LO state. It was shown in \cite{Buballa2009} that
the self-consistent 1D modulation forms a soliton lattice structure near the lower critical field, which lowers the grand potential of the system. Near the upper critical field the gap function becomes sinusoidal, and therefore the grand potentials of the LO state and the 1D modulation agree with each other. We notice that the phase transition from the LO state to the normal state is of first order, which is in contradiction to the prediction from the GL analysis. To understand the reason, we show in Fig. \ref{LO} (b) the potential curve at $\delta\mu=0.775\Delta_0$ and at the optimal pair momentum $q=1.1613\delta\mu$. We find that the potential curve has two minima: a shallow minimum at $\Delta\simeq0.12\Delta_0$ and a deep minimum at $\Delta\simeq0.44\Delta_0$. Obviously, the shallow minimum is responsible for the GL theory which predicts a second-order phase transition. However, the deep global minimum, which cannot be captured by the GL theory up to the order $\Delta^6$, is responsible for the real first-order phase transition. Therefore, the LO state already shows the importance of the higher-order expansions in the GL theory.

\section{Calculation of the Grand Potential: Small Block Method}

The key problem in the numerical calculation is the diagonalization of the matrix ${\bf H}_+$ or ${\bf H}_-$ and obtaining all the eigenvalues. For BCC and FCC structures, we use a symmetrical truncation $-D<l,m,n<D$ with a large cutoff $D\in \mathbb{Z}^+$. However, the matrix size grows dramatically with increasing cutoff $D$, which makes the calculation infeasible because of not only the computing capability of current computing facilities but also the computing time and cost. Notice that we need to diagonalize the matrix ${\bf H}_+$ for various values of the momentum ${\bf k}$ in the BZ, the gap parameter $\Delta$, and the pair momentum $q$.

We first estimate the size of $D$ needed for the convergence of the grand potential $\delta\Omega$. The matrix size $(2D+1)^3$ and hence the computing time and cost grow dramatically with increasing $D$. The cutoff $D$ is related to the maximum momentum $k_{\rm max}$ in each direction ($x$, $y$, and $z$). We have
\begin{eqnarray}
k_{\rm max}=(2D+1)\frac{\pi}{a}.
\end{eqnarray}
This maximum momentum can be roughly estimated from the calculation of the LO state. For the LO state, the matrix size becomes $2D+1$ and exact diagonalization is possible. The regime of $\delta\mu$ we are interested in is $\delta\mu/\Delta_0\in[0.7-0.8]$ and the optimal pair momentum is located at $q\simeq\delta\mu$. From the calculation of the LO state at $\Delta_0\sim100$ MeV, we find that $k_{\rm max}$ must reach at least $5$GeV for convergence. Notice that we have $k_{\rm max}=(2D+1)q$ for BCC and $\sqrt{3}k_{\rm max}=(2D+1)q$ for FCC.  Therefore, the cutoff $D$ for BCC can be estimated as $D\sim35$, which corresponds to a matrix size $\sim3\times10^5$. For FCC, the cutoff is even larger because of the factor $\sqrt{3}$. We have $D\sim60$ for FCC, which corresponds to a matrix size $\sim1.5\times10^6$. Notice that this is only a naive estimation. In practice, the cutoff needed for convergence may be smaller or larger. Exact diagonalization of such huge matrices to obtain all the eigenvalues are impossible with our current computing facility.

We therefore need a feasible scheme to evaluate the grand potential $\delta\Omega$. Notice that decreasing the value of $\Delta_0$ does not reduce the size of the matrices. In this case, even though $k_{\rm max}$ becomes smaller, the pair momentum $q$ also gets smaller. Let us call an off-diagonal element $\Delta$ in ${\bf H}_+$ or ${\bf H}_-$ a ``coupling". Because our goal is to evaluate the grand potential $\delta\Omega$ rather than to know
exactly all the band dispersions (eigenvalues), we may neglect a small amount of couplings to lower the size of the matrices. By neglecting this small amount of couplings, the huge matrix ${\bf H}_+$ becomes block diagonal with each block having a much smaller size. In general, we expect that the omission of a small amount of couplings $\Delta$ induces only a perturbation to the grand potential $\delta\Omega$. We shall call this scheme small block method (SBM).

\begin{figure*}[!htb]
\begin{center}
\includegraphics[width=8.4cm]{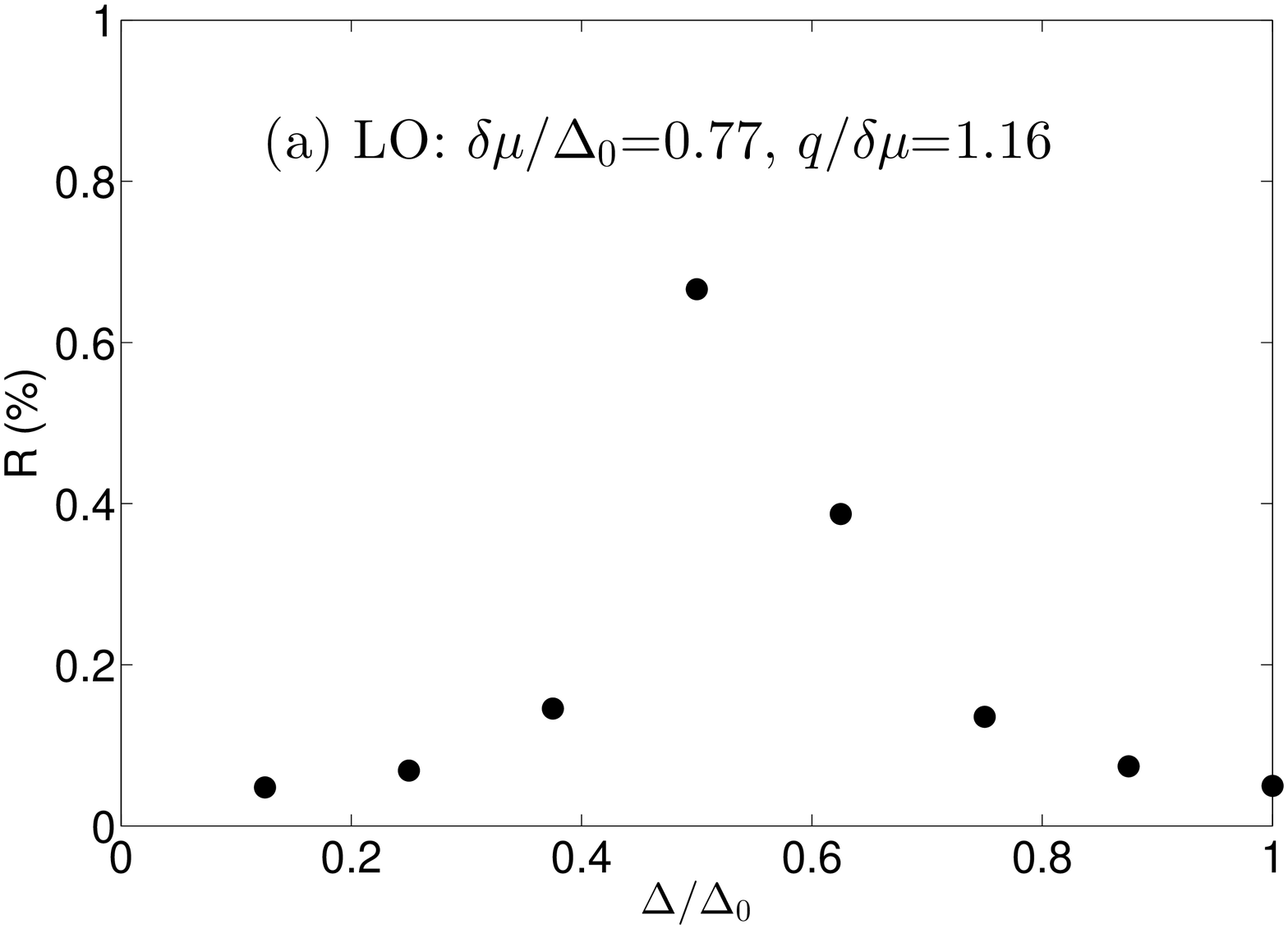}
\includegraphics[width=8.4cm]{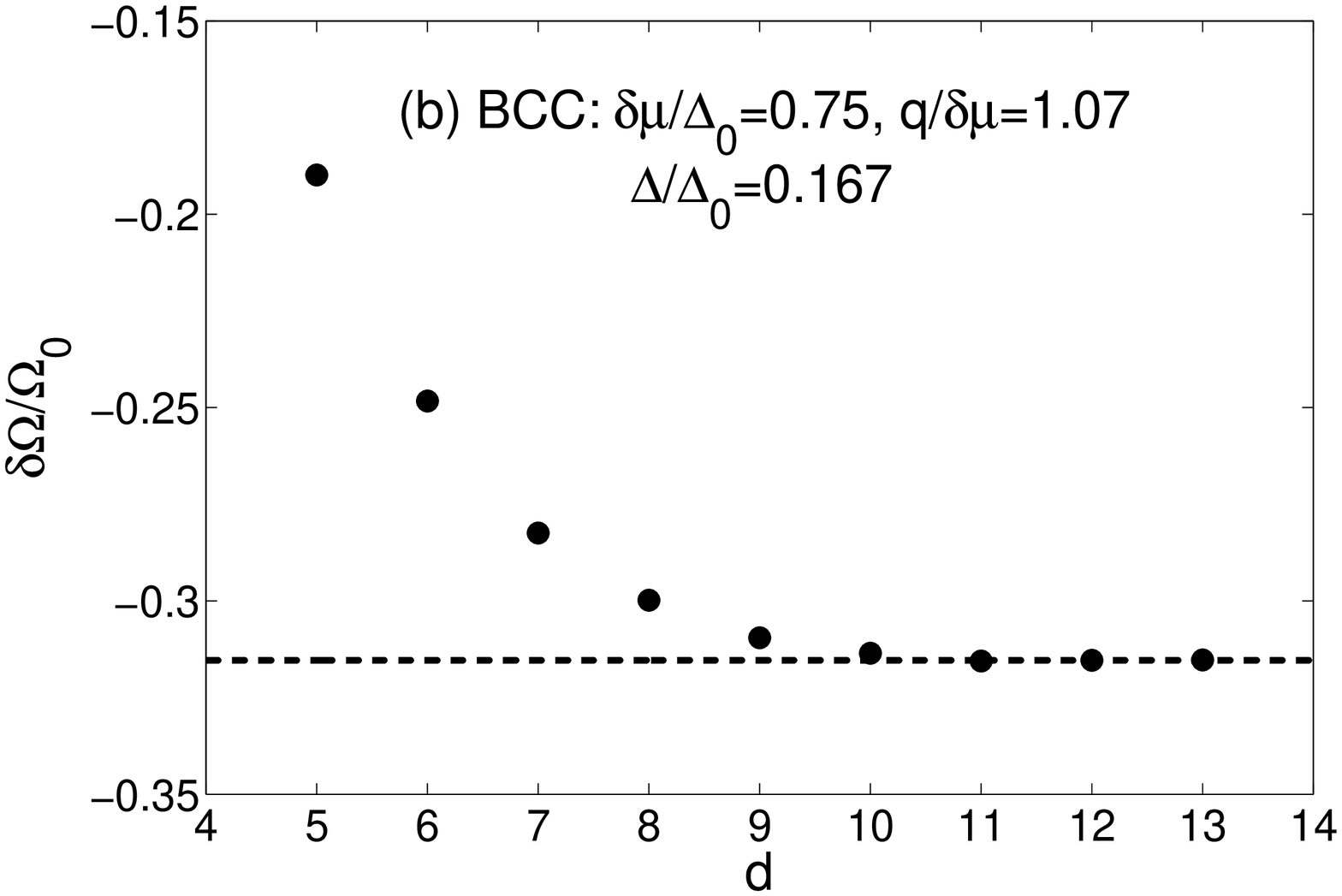}
\caption{Comparison of the grand potentials calculated from the exact diagonalization and from the small block method. (a) The relative error $R$ for the LO state at $\delta\mu/\Delta_0=0.77$ and $q/\delta\mu=1.16$ with $D=50$ and $d=20$. (b) The grand potential for the BCC state at $\delta\mu/\Delta_0=0.75$, $q/\delta\mu=1.07$, and $\Delta/\Delta_0=0.167$ as a function of $d$.  \label{ERROR}}
\end{center}
\end{figure*}

To be specific, the size of the small blocks in our calculation is $(2d+1)^3$ with $d\in\mathbb{Z}^+$. In general, we have $d<D$. For symmetry reason, we require that the centers of these blocks are located at the reciprocal lattice vectors
\begin{eqnarray}
{\bf G}_{[n_x,n_y,n_z]}=(2d+1)\frac{2\pi}{a}(n_x{\bf e}_x+n_y{\bf e}_y+n_z{\bf e}_z)
\end{eqnarray}
with $n_x,n_y,n_z\in\mathbb{Z}$. This scheme makes the SBM feasible even though $(2D+1)^3$ is not divisible by $(2d+1)^3$.
In practice, we first choose a large cutoff $D$ which is sufficient for convergence. By increasing the value of $d$, we find that
the grand potential $\delta\Omega$ finally converges. In practice, if the grand potentials evaluated at several values of $d$, i.e., $d_0-k$, $d_0-k+1$, ... , and $d_0$ ($k\in\mathbb{Z}^+$), are very close to each other, we identify that the grand potential converges to its precise value from exact diagonalization.  At the converging value $d=d_0$, the block size $(2d_0+1)^3$ is normally much smaller than the total size $(2D+1)^3$. This scheme makes the calculation feasible and also saves a lot of computing time and cost.

The matrices for the 3D structures are huge and cannot be written down here. For the sake of simplicity, let us use the LO state as a toy example for the SBM. In this case, the matrix size and the block size are $2D+1$ and $2d+1$, respectively. The centers of the blocks are located at the reciprocal lattice vectors $(2d+1)2qn_z{\bf e}_z$ with $n_z\in\mathbb{Z}$. For $D=10$, the matrix ${\bf H}_+$ reads
\begin{widetext}
\begin{eqnarray}
\left(\begin{array}{ccccccccccccccccccccc}
\varepsilon_{+10} & \Delta & 0 & 0 & 0 & 0 & 0 & 0 & 0 & 0 & 0 & 0 & 0 & 0 & 0 & 0 & 0 & 0 & 0 & 0 & 0\\
\Delta & \varepsilon_{+9} & \Delta & 0 & 0 & 0 & 0 & 0 & 0 & 0 & 0 & 0 & 0 & 0 & 0 & 0 & 0 & 0 & 0 & 0 & 0\\
0 & \Delta & \varepsilon_{+8} & \color{red}{\bf\Delta} & 0 & 0 & 0 & 0 & 0 & 0 & 0 & 0 & 0 & 0 & 0 & 0 & 0 & 0 & 0 & 0 & 0\\
0 & 0 & \color{red}{\bf \Delta} & \varepsilon_{+7} & \Delta & 0 & 0 & 0 & 0 & 0 & 0 & 0 & 0 & 0 & 0 & 0 & 0 & 0 & 0 & 0 & 0\\
0 & 0 & 0 & \Delta & \varepsilon_{+6} & \Delta & 0 & 0 & 0 & 0 & 0 & 0 & 0 & 0 & 0 & 0 & 0 & 0 & 0 & 0 & 0\\
0 & 0 & 0 & 0 & \Delta & \varepsilon_{+5} & \Delta & 0 & 0 & 0 & 0 & 0 & 0 & 0 & 0 & 0 & 0 & 0 & 0 & 0 & 0\\
0 & 0 & 0 & 0 & 0 & \Delta & \varepsilon_{+4} & \Delta & 0 & 0 & 0 & 0 & 0 & 0 & 0 & 0 & 0 & 0 & 0 & 0 & 0\\
0 & 0 & 0 & 0 & 0 & 0 & \Delta & \varepsilon_{+3} & \color{red}{\bf\Delta} & 0 & 0 & 0 & 0 & 0 & 0 & 0 & 0 & 0 & 0 & 0 & 0\\
0 & 0 & 0 & 0 & 0 & 0 & 0 & \color{red}{\bf\Delta} & \varepsilon_{+2} & \Delta & 0 & 0 & 0 & 0 & 0 & 0 & 0 & 0 & 0 & 0 & 0\\
0 & 0 & 0 & 0 & 0 & 0 & 0 & 0 & \Delta & \varepsilon_{+1} & \Delta & 0 & 0 & 0 & 0 & 0 & 0 & 0 & 0 & 0 & 0\\
0 & 0 & 0 & 0 & 0 & 0 & 0 & 0 & 0 & \Delta & \varepsilon_0 & \Delta & 0 & 0 & 0 & 0 & 0 & 0 & 0 & 0 & 0\\
0 & 0 & 0 & 0 & 0 & 0 & 0 & 0 & 0 & 0 & \Delta & \varepsilon_{-1} & \Delta & 0 & 0 & 0 & 0 & 0 & 0 & 0 & 0\\
0 & 0 & 0 & 0 & 0 & 0 & 0 & 0 & 0 & 0 & 0 & \Delta & \varepsilon_{-2} & \color{red}{\bf\Delta} & 0 & 0 & 0 & 0 & 0 & 0 & 0\\
0 & 0 & 0 & 0 & 0 & 0 & 0 & 0 & 0 & 0 & 0 & 0 & \color{red}{\bf\Delta} & \varepsilon_{-3} & \Delta & 0 & 0 & 0 & 0 & 0 & 0\\
0 & 0 & 0 & 0 & 0 & 0 & 0 & 0 & 0 & 0 & 0 & 0 & 0 & \Delta & \varepsilon_{-4} & \Delta & 0 & 0 & 0 & 0 & 0\\
0 & 0 & 0 & 0 & 0 & 0 & 0 & 0 & 0 & 0 & 0 & 0 & 0 & 0 & \Delta & \varepsilon_{-5} & \Delta & 0 & 0 & 0 & 0\\
0 & 0 & 0 & 0 & 0 & 0 & 0 & 0 & 0 & 0 & 0 & 0 & 0 & 0 & 0 & \Delta & \varepsilon_{-6} & \Delta & 0 & 0 & 0\\
0 & 0 & 0 & 0 & 0 & 0 & 0 & 0 & 0 & 0 & 0 & 0 & 0 & 0 & 0 & 0 & \Delta & \varepsilon_{-7} & \color{red}{\bf\Delta} & 0 & 0\\
0 & 0 & 0 & 0 & 0 & 0 & 0 & 0 & 0 & 0 & 0 & 0 & 0 & 0 & 0 & 0 & 0 & \color{red}{\bf\Delta} & \varepsilon_{-8} & \Delta & 0\\
0 & 0 & 0 & 0 & 0 & 0 & 0 & 0 & 0 & 0 & 0 & 0 & 0 & 0 & 0 & 0 & 0 & 0 & \Delta & \varepsilon_{-9} & \Delta\\
0 & 0 & 0 & 0 & 0 & 0 & 0 & 0 & 0 & 0 & 0 & 0 & 0 & 0 & 0 & 0 & 0 & 0 & 0 & \Delta & \varepsilon_{-10}
\end{array}
\right),
\end{eqnarray}
where $\varepsilon_n=(-1)^n\left[\sqrt{{\bf k}_\perp^2 + (k_z +2nq)^2}- \mu\right]$. If we take $d=2$, we neglect the couplings $\Delta$ in red. In this case, the matrix ${\bf H}_+$ is approximated as
\begin{eqnarray}
\left(
\begin{array}{ccccccccccccccccccccc}
\varepsilon_{+10} & \Delta & 0 & 0 & 0 & 0 & 0 & 0 & 0 & 0 & 0 & 0 & 0 & 0 & 0 & 0 & 0 & 0 & 0 & 0 & 0\\
\Delta & \varepsilon_{+9} & \Delta & 0 & 0 & 0 & 0 & 0 & 0 & 0 & 0 & 0 & 0 & 0 & 0 & 0 & 0 & 0 & 0 & 0 & 0\\
0 & \Delta & \varepsilon_{+8} & \color{red}{\bf0} & 0 & 0 & 0 & 0 & 0 & 0 & 0 & 0 & 0 & 0 & 0 & 0 & 0 & 0 & 0 & 0 & 0\\
0 & 0 & \color{red}{\bf0} & \varepsilon_{+7} & \Delta & 0 & 0 & 0 & 0 & 0 & 0 & 0 & 0 & 0 & 0 & 0 & 0 & 0 & 0 & 0 & 0\\
0 & 0 & 0 & \Delta & \varepsilon_{+6} & \Delta & 0 & 0 & 0 & 0 & 0 & 0 & 0 & 0 & 0 & 0 & 0 & 0 & 0 & 0 & 0\\
0 & 0 & 0 & 0 & \Delta & \varepsilon_{+5} & \Delta & 0 & 0 & 0 & 0 & 0 & 0 & 0 & 0 & 0 & 0 & 0 & 0 & 0 & 0\\
0 & 0 & 0 & 0 & 0 & \Delta & \varepsilon_{+4} & \Delta & 0 & 0 & 0 & 0 & 0 & 0 & 0 & 0 & 0 & 0 & 0 & 0 & 0\\
0 & 0 & 0 & 0 & 0 & 0 & \Delta & \varepsilon_{+3} & \color{red}{\bf0} & 0 & 0 & 0 & 0 & 0 & 0 & 0 & 0 & 0 & 0 & 0 & 0\\
0 & 0 & 0 & 0 & 0 & 0 & 0 & \color{red}{\bf0} & \varepsilon_{+2} & \Delta & 0 & 0 & 0 & 0 & 0 & 0 & 0 & 0 & 0 & 0 & 0\\
0 & 0 & 0 & 0 & 0 & 0 & 0 & 0 & \Delta & \varepsilon_{+1} & \Delta & 0 & 0 & 0 & 0 & 0 & 0 & 0 & 0 & 0 & 0\\
0 & 0 & 0 & 0 & 0 & 0 & 0 & 0 & 0 & \Delta & \varepsilon_0 & \Delta & 0 & 0 & 0 & 0 & 0 & 0 & 0 & 0 & 0\\
0 & 0 & 0 & 0 & 0 & 0 & 0 & 0 & 0 & 0 & \Delta & \varepsilon_{-1} & \Delta & 0 & 0 & 0 & 0 & 0 & 0 & 0 & 0\\
0 & 0 & 0 & 0 & 0 & 0 & 0 & 0 & 0 & 0 & 0 & \Delta & \varepsilon_{-2} & \color{red}{\bf0} & 0 & 0 & 0 & 0 & 0 & 0 & 0\\
0 & 0 & 0 & 0 & 0 & 0 & 0 & 0 & 0 & 0 & 0 & 0 & \color{red}{\bf0} & \varepsilon_{-3} & \Delta & 0 & 0 & 0 & 0 & 0 & 0\\
0 & 0 & 0 & 0 & 0 & 0 & 0 & 0 & 0 & 0 & 0 & 0 & 0 & \Delta & \varepsilon_{-4} & \Delta & 0 & 0 & 0 & 0 & 0\\
0 & 0 & 0 & 0 & 0 & 0 & 0 & 0 & 0 & 0 & 0 & 0 & 0 & 0 & \Delta & \varepsilon_{-5} & \Delta & 0 & 0 & 0 & 0\\
0 & 0 & 0 & 0 & 0 & 0 & 0 & 0 & 0 & 0 & 0 & 0 & 0 & 0 & 0 & \Delta & \varepsilon_{-6} & \Delta & 0 & 0 & 0\\
0 & 0 & 0 & 0 & 0 & 0 & 0 & 0 & 0 & 0 & 0 & 0 & 0 & 0 & 0 & 0 & \Delta & \varepsilon_{-7} & \color{red}{\bf0} & 0 & 0\\
0 & 0 & 0 & 0 & 0 & 0 & 0 & 0 & 0 & 0 & 0 & 0 & 0 & 0 & 0 & 0 & 0 & \color{red}{\bf0} & \varepsilon_{-8} & \Delta & 0\\
0 & 0 & 0 & 0 & 0 & 0 & 0 & 0 & 0 & 0 & 0 & 0 & 0 & 0 & 0 & 0 & 0 & 0 & \Delta & \varepsilon_{-9} & \Delta\\
0 & 0 & 0 & 0 & 0 & 0 & 0 & 0 & 0 & 0 & 0 & 0 & 0 & 0 & 0 & 0 & 0 & 0 & 0 & \Delta & \varepsilon_{-10}
\end{array}
\right).
\end{eqnarray}
\end{widetext}
Therefore, by neglecting a small amount of couplings, we have made the large matrix ${\bf H}_+$ block-diagonal. Notice that this is only a toy example for the SBM. In practice, $D=10$ and $d=2$ is obviously not enough for convergence.

For the LO state, exact diagonalization of the matrices at $q\simeq\delta\mu$ is quite easy because the size of the matrices is $2D+1$. We can therefore check the error induced by the SBM. The relative error induced by the SBM can be defined as
\begin{eqnarray}
R=\frac{|\delta\Omega_{\rm SBM}-\delta\Omega_{\rm EX}|}{\delta\Omega_{\rm EX}},
\end{eqnarray}
where $\delta\Omega_{\rm SBM}$ and $\delta\Omega_{\rm EX}$ are the grand potentials obtained from the SBM and exact diagonalization, respectively.
In Fig. \ref{ERROR} (a), we show a numerical example of the relative error for the LO state at $\delta\mu/\Delta_0=0.77$ and $q/\delta\mu=1.16$. In the calculations, we use $D=50$ and $d=20$. We find that the relative error is very small, generally of order $O(10^{-3})$. The slightly larger error around $\Delta/\Delta_0=0.5$ is due to the fact that $\delta\Omega$ itself is very small there. For the BCC structure, we are not able to check the relative error at $q\simeq\delta\mu$ because it is impossible to exactly diagonalize the matrices with a huge size $(2D+1)^3$. However, we can check the $d$ dependence of the grand potential. For pair momentum around $q\simeq\delta\mu$, we choose a sufficiently large cutoff $D$ and increase the value of $d$. We evaluate the grand potentials for various values of $d$ (i.e., $d_0-k$, $d_0-k+1$, ... , and $d_0$). If they are very close to each other, we identify that the grand potential converges. Then the grand potential $\delta\Omega$ can be evaluated at $d=d_0$. In Fig. \ref{ERROR}, we show the $d$ dependence of the grand potential of the BCC structure at $\delta\mu/\Delta_0=0.75$, $q/\delta\mu=1.07$, and $\Delta/\Delta_0=0.167$. In the calculation we choose $D=50$ which is sufficiently large to guarantee the convergence at large ${\bf G}$. We find that for BCC structure, $d_0$ is normally within the range $10<d_0<15$, which is feasible for a calculation. For FCC structure, we do not find a satisfying convergence at these small values of $d$.


\begin{thebibliography}{99}
\bibitem{note1}           {The grand potential $\Omega(\Delta,q)$ is still logarithmically divergent after applying the Pauli-Villars regularization.
                           However, by subtracting $\Omega(0,q)$ for the normal state, $\delta\Omega(\Delta,q)$ is well defined.}
\bibitem{note2}           {For very small values of $\Delta_0$ (such as $\Delta_0=20$ MeV), the grand potential $\delta\Omega$ becomes also very
                           small. We therefore need much higher accuary in diagonalizing the matrices, which leads to much larger computing time
                           and cost.}
\bibitem{note4}           {Note that decreasing the value of $\Delta_0$ does not help to reduce the matrix size. With decreasing $\Delta_0$,
                           even though $k_{\rm max}$ becomes smaller, the optimal pair momentum also gets smaller.}
\bibitem{note5}           {The parameter $\Lambda$ for the Pauli-Villars regularization is chosen as $\Lambda=400$ MeV. However, we checked
                           that a different choice of $\Lambda$ does not leads to a qualitative different result.}
\end{thebibliography}
\end{document}